\documentclass[12pt,a4paper,floatfix]{iopart}

\usepackage{bm}
\usepackage{graphicx}
\usepackage{amssymb}
\usepackage{latexsym}
\usepackage{psfrag}
\usepackage{color}
\usepackage{textcomp}

\def\refs{\marginpar {\scriptsize REFERENCES}}

\def\beq{\begin{equation}}
\def\eeq{\end{equation}} 
\def\bea{\begin{eqnarray}}
\def\eea{\end{eqnarray}}
\def\benu{\begin{enumerate}}
\def\eenu{\end{enumerate}}
\def\nn{\nonumber} 

\def\f{\frac}
\def\l{\left}
\def\r{\right}
\def\d{{\rm d}}
\def\cR{{\mathcal{R}}}

\def\ei{\eta_{\mathrm{i}}}
\def\es{\eta_{\mathrm{s}}}
\def\ee{\eta_{\mathrm{e}}}

\def\Ns{N_{\mathrm{s}}}
\def\Ne{N_{\mathrm{e}}}

\def\vx{{\bm x}}
\def\vk{{\bm k}}
\def\vka{{\bm k}_{1}}
\def\vkb{{\bm k}_{2}}
\def\vkc{{\bm k}_{3}}
\def\ka{k_{1}}
\def\kb{k_{2}}
\def\kc{k_{3}}
\def\cB{{\cal B}}
\def\cG{{\cal G}}
\def\fnl{f_{_{\rm NL}}}
\def\Mpl{M_{_{\rm Pl}}}

\newcommand{\uPl}{\mathrm{Pl}}

\newcommand{\usssPl}{\sss{\uPl}}

\newcommand{\Mp}{M_\usssPl}

\newcommand{\email}[1]{\ead{#1}}
\newcommand{\affiliation}[1]{\address{#1}}

\newcommand{\sss}[1]{{\scriptscriptstyle{#1}}}

\begin{document}

\title{BINGO: A code for the efficient computation of the scalar 
bi-spectrum}
\author{Dhiraj Kumar Hazra}
\affiliation{Harish-Chandra Research Institute, Chhatnag Road,
Jhunsi, Allahabad~211019, India\footnote{Current affiliation:~Asia 
Pacific Center for Theoretical Physics, Pohang, Gyeongbuk 
790-784, Korea. E-mail:~{\tt dhiraj@apctp.org}.}.}
\author{L.~Sriramkumar}
\affiliation{Department of Physics, Indian Institute of Technology 
Madras, Chennai~600036, India.} 
\email{sriram@physics.iitm.ac.in}
\author{J\'er\^ome Martin}
\affiliation{Institut d'Astrophysique de Paris, UMR7095-CNRS, 
Universit\'e Pierre et Marie Curie, 98bis boulevard Arago,
75014 Paris, France.}
\email{jmartin@iap.fr}
\date{\today}
\begin{abstract}
  We present a new and accurate Fortran code, the BI-spectra and Non-Gaussianity
  Operator (BINGO), for the efficient numerical computation of the scalar
  bi-spectrum and the non-Gaussianity parameter $\fnl$ in single field
  inflationary models involving the canonical scalar field. The code can calculate
  all the different contributions to the bi-spectrum and the parameter $\fnl$ for
  an arbitrary triangular configuration of the wavevectors. Focusing firstly on the
  equilateral limit, we illustrate the accuracy of BINGO by comparing the
  results from the code with the spectral dependence of the bi-spectrum expected
  in power law inflation. Then, considering an arbitrary triangular configuration,
  we contrast the numerical results with the analytical expression available in
  the slow roll limit, for, say, the case of the conventional quadratic potential.
  Considering a non-trivial scenario involving deviations from slow roll, we
  compare the results from the code with the analytical results that have recently
  been obtained in the case of the Starobinsky model in the equilateral limit. As
  an immediate application, we utilize BINGO to examine of the power of the
  non-Gaussianity parameter $\fnl$ to discriminate between various inflationary
  models that admit departures from slow roll and lead to similar features in the
  scalar power spectrum. We close with a summary and discussion on the implications 
  of the results we obtain.
\end{abstract}
\maketitle
\flushbottom




\section{Non-Gaussianities and primordial features}

Over the last half-a-dozen years, it has been increasingly realized 
that the detection of non-Gaussianities in the primordial perturbations 
can considerably help in constraining the inflationary 
models (see Refs.~\cite{maldacena-2003,ng-ncsf,ng-reviews}; for early 
efforts in this direction, see Refs.~\cite{earlyng}).
In particular, the detection of a high value for the dimensionless 
non-Gaussianity parameter $\fnl$ that is often used to describe the 
amplitude of the scalar bi-spectrum can rule out a wide class of models.
For instance, if the extent of non-Gaussianity actually proves to be as 
large as the mean values of $\fnl$ arrived at from the WMAP 
data (see Refs.~\cite{wmap-2009,wmap-2011,wmap-2012}; in this context, 
also see Refs.~\cite{ng-da,ng-da-reviews}), then canonical scalar field 
models that lead to slow roll inflation and nearly scale invariant 
primordial spectra will cease to be consistent with the data. 

\par

However, the evaluation of the scalar bi-spectrum in a generic inflationary 
model proves to be a non-trivial task~\cite{funakoshi-2012}.
Often, it is the slow roll approximation that is resorted to in order to 
arrive at analytical expressions for the scalar bi-spectrum and the 
non-Gaussianity parameter~$\fnl$~\cite{maldacena-2003,ng-ncsf}.
While the slow roll approximation can actually encompass a relatively wide 
class of models, evidently, the approach will cease to be valid when 
departures from slow roll occur.
In such situations, one is often forced to approach the problem numerically.  
In this work, we shall present a new and accurate Fortran code, the
BI-spectra and Non-Gaussianity Operator (BINGO), to numerically 
evaluate the scalar bi-spectrum and the non-Gaussianity parameter~$\fnl$ for 
single field inflationary models involving the canonical scalar field. 
Although, some partial numerical results have already been published in the 
literature, we believe that it is for the first time that a general 
and efficient code has been put together to evaluate the complete scalar 
bi-spectrum. 
Though, we shall largely restrict our discussion in this paper to the 
equilateral case, as we shall illustrate, the code can compute the 
scalar bi-spectrum and the parameter~$\fnl$ for any triangular 
configuration of the wavevectors.
Also, as we shall demonstrate, BINGO can also compute {\it all}\/ the different 
contributions to the bi-spectrum.
Moreover, it is worth noting that, under certain conditions, the code can 
arrive at the results within a matter of a few minutes.
We should mention here that we have made a limited version of BINGO, viz.
one that focuses on the equilateral limit, available online at 
{\tt https://www.physics.iitm.ac.in/{\textasciitilde{}}sriram/bingo/bingo.html}.

\par

As an immediate application, we shall utilize the code to examine of the 
power of the non-Gaussianity parameter $\fnl$ to discriminate between 
different inflationary models that admit deviations from slow roll and
generate similar features in the scalar power spectrum.
Recall that, most single field inflationary models naturally lead to an 
extended period of slow roll and hence to nearly scale invariant primordial 
power spectra (see, for example, any of the following texts~\cite{texts} 
or reviews~\cite{reviews}), which seem to be fairly consistent with 
the recent data on the anisotropies in the Cosmic Microwave Background 
(CMB) (in this context, see Refs.~\cite{wmap-2009,wmap-2011,wmap-2012};
for a comparison of a class of inflationary models with the data, see, for 
example, Ref.~\cite{martin-2011} and references therein) as well as 
other observational constraints.
However, it has been repeatedly noticed that certain features in the 
inflationary scalar power spectrum can improve the fit to the CMB data
at the cost of some additional parameters (see, for instance, 
Refs.~\cite{wmap-2003,rc,quadrupole,pi,l2240,hazra-2010,benetti-2011,joy-2008-2009,general,quadrupole-sic,pso,pahud-2009,flauger-2010-2011,kobayashi-2011,aich-2011}).
Though the statistical significance of these features remain to be 
understood satisfactorily~\cite{rc-wf}, they gain importance from the 
phenomenological perspective of comparing the models with the data, 
because only a smaller class of single field inflationary models, 
which allow for departures from slow roll, can generate them.
Interestingly, demanding the presence of features in the scalar 
power spectrum seems to generically lead to larger non-Gaussianities 
(see, for example, Refs.~\cite{ng-f1,ng-f}).
Therefore, features may offer the only route (unless one works with  
non-vacuum initial states~\cite{ng-nvis}) for the canonical 
scalar fields to remain viable if $\fnl$ turns out to be 
significant.

\par

If indeed the presence of features turns out to be the correct reason
behind possibly large non-Gaussianities, can we observationally identify 
the correct underlying inflationary scenario, in particular, given the 
fact that different models can lead to similar features in the scalar 
power spectrum? In other words, to what extent can the non-Gaussianity 
parameter $\fnl$ help us discriminate between the inflationary models 
that permit features? To address this question, we shall consider a
few typical inflationary models leading to features, assuming that they 
can be viewed as representatives of such a class of scenarios. Concretely, 
we shall consider the Starobinsky model~\cite{starobinsky-1992} and the 
punctuated inflationary scenario~\cite{pi}, both of which result in a 
sharp drop in power at large scales that is followed by oscillations.  
We shall also study large and small field models with an additional step 
introduced in the inflaton potential~\cite{hazra-2010,ng-f1,ng-f}. The 
step leads to a burst of oscillations in the scalar power spectrum which 
improve the fit to the outliers near the multipole moments of $\ell=22$ 
and $40$ in the CMB anisotropies. We shall also consider oscillating 
inflaton potentials such as the one that arises in the axion monodromy 
model which lead to modulations in the power spectrum over a wide range of
scales~\cite{pahud-2009,flauger-2010-2011,kobayashi-2011,aich-2011}.

\par

The plan of this paper is as follows.  
In the following section, we shall quickly describe a few essential 
details pertaining to the power spectrum and the bi-spectrum as well 
as the non-Gaussianity parameter $\fnl$ in inflationary models 
involving a single, canonical, scalar field. 
In Sec.~\ref{sec:nc-sbs}, after demonstrating that the super Hubble 
contributions to the bi-spectrum during inflation prove to be negligible, 
we shall describe the method that BINGO adopts to numerically compute the 
bi-spectrum and the non-Gaussianity parameter $\fnl$.
We shall also illustrate the extent of accuracy of BINGO by comparing the 
numerical results with the analytical results available in three situations.
Firstly, restricting ourselves to the equilateral limit, we shall compare 
the results from the numerical computations with the spectral dependence 
of the bi-spectrum expected in power law inflation.
Secondly, focusing on the case of the archetypical quadratic potential, 
we shall contrast the numerical results with the analytical expressions 
obtained using the slow roll approximation for an arbitrary triangular 
configuration of the wavevectors. 
Thirdly, considering a more non-trivial situation involving a brief period 
of fast roll, we shall compare the numerical results with the analytical 
results that have recently been obtained in the case of the Starobinsky 
model in the equilateral limit~\cite{martin-2012a,arroja-2011-2012}. 
In the succeeding two sections, we shall utilize BINGO to examine of the power 
of the non-Gaussianity parameter $\fnl$ to discriminate between different 
inflationary models that permit deviations from slow roll and generate 
similar features in the scalar power spectrum.
After a quick outline of the inflationary models of our interest, we shall 
discuss the scalar power spectra that arise in these models.
We shall then present the main results, and compare the $\fnl$ that arise 
in the various models that we consider.
We shall finally conclude in Sec.~\ref{sec:d} with a brief summary and
outlook.

\par

A few words on our conventions and notations are in order before we 
proceed.  
We shall work with units such that $c=\hbar=1$, and we shall set $\Mp^2 
= (8\, \pi\, G)^{-1}$.  
As is often done in the context of inflation, we shall assume the 
background to be described by the spatially flat, 
Friedmann-Lemaitre-Robertson-Walker line-element.
We shall denote the conformal time as $\eta$, while $N$ shall represent 
the number of e-folds.
Moreover, an overprime shall denote differentiation with respect to the 
conformal time coordinate $\eta$.
Lastly, $a$ shall denote the scale factor, with the corresponding Hubble 
parameter being defined as $H=a'/a^2$.


\section{The scalar power spectrum, bi-spectrum and the 
parameter $\fnl$}\label{subsec:dbs-dc}

In this section, we shall first rapidly summarize the essential definitions
of the power spectrum and the bi-spectrum as well as the non-Gaussianity 
parameter $\fnl$.
We shall then list the various contributions to the bi-spectrum that arise 
in the Maldacena formalism (for the original discussions, see 
Refs.~\cite{maldacena-2003,ng-ncsf,ng-reviews,ng-f1,ng-f}; for more recent 
discussions, see, for example, Refs.~\cite{martin-2012a,hazra-2012}).


\subsection{Essential definitions and relations}

Let us begin by recalling some essential points concerning the power
spectrum.  On quantization, the operator corresponding to the
curvature perturbation $\cR(\eta ,\vx)$ can be expressed as
\begin{eqnarray}
{\hat \cR}(\eta ,\vx)
&=&\int \f{{\rm d}^{3}{\vk}}{(2\,\pi)^{3/2}}\, 
{\hat {\cal R}}_{\vk}(\eta)\; {\rm e}^{i\, \vk \cdot \vx}\nn\\
&=&\int \f{\d^{3}{\vk}}{(2\,\pi)^{3/2}}\, 
\l[{\hat a}_{\vk}\, f_{\vk}(\eta)\, {\rm e}^{i\,\vk\cdot \vx}
+{\hat a}_{\vk}^{\dagger}\, f_\vk^*(\eta)\, 
{\rm e}^{-i\,\vk \cdot \vx}\r],\label{eq:cR-d}
\end{eqnarray}
where ${\hat a}_{\vk}$ and ${\hat a}_{\vk}^{\dagger}$ are the usual 
creation and annihilation operators that satisfy the standard 
commutation relations.
The modes $f_\vk$ are governed by the differential 
equation~\cite{texts,reviews}
\begin{equation}
f_\vk''+2\, \f{z'}{z}\, f_\vk' + k^{2}\, f_{\vk}=0,\label{eq:de-fk}
\end{equation}
where $z=a\,\Mp\,\sqrt{2\,\epsilon_1}$, with $\epsilon_1=
-\d\, {\rm ln}\,H/\d N$ being the first slow roll parameter.
The dimensionless scalar power spectrum ${\cal P}_{_{\rm S}}(k)$ 
is defined in terms of the correlation function of the Fourier 
modes of the curvature perturbation ${\hat {\cal R}}_{\vk}$ as 
follows:
\begin{equation}
\label{eq:ps}
\langle 0\vert {\hat \cR}_{\vk}(\eta)\, 
{\hat \cR}_{\bf p}(\eta)\vert 0\rangle 
=\f{(2\, \pi)^2}{2\, k^3}\; {\cal P}_{_{\rm S}}(k)\;
\delta^{(3)}\l(\vk+{\bm p}\r),
\end{equation}
where $\vert 0\rangle$ denotes the Bunch-Davies vacuum which 
is defined as ${\hat a}_{\vk}\vert 0 \rangle=0$ $\forall$ 
$\vk$~\cite{bunch-1978}.
In terms of the modes $f_\vk$, the scalar power spectrum is given 
by
\begin{equation}
{\cal P}_{_{\mathrm{S}}}(k)
=\frac{k^{3}}{2\, \pi^{2}}\, \vert f_{\vk}\vert^{2}.
\label{eq:sps-d}
\end{equation}
As is well known, numerically, the initial conditions are imposed on
the modes $f_\vk$ when they are well inside the Hubble radius, and 
the power spectrum is evaluated at suitably late times when the modes 
are sufficiently outside the Hubble radius (see, for instance, 
Refs.~\cite{ne-ps}).  
We shall discuss the details concerning the numerical evolution of the 
background as well as the perturbations and the computation of the 
corresponding power spectrum in the next section.

\par

The scalar bi-spectrum $\cB_{_{\rm S}}(\vka,\vkb,\vkc)$ is related 
to the three point correlation function of the Fourier modes of the 
curvature perturbation ${\hat \cR}_{\bm k}$, evaluated at the end of 
inflation, say, at the conformal time $\eta_{\rm e}$, as 
follows~\cite{wmap-2011}: 
\begin{equation}
\langle {\hat \cR}_{\vka}(\ee)\, {\hat \cR}_{\vkb}(\ee)\, 
{\hat \cR}_{\vkc}(\ee)\rangle 
=\l(2\,\pi\r)^3\, \cB_{_{\rm S}}(\vka,\vkb,\vkc)\;
\delta^{(3)}\l(\vka+\vkb+\vkc\r).\label{eq:bi-s}
\end{equation}
In our discussion below, for the sake of convenience, we shall set
\begin{equation}
\cB_{_{\mathrm{S}}}(\vka,\vkb,\vkc)=\l(2\,\pi\r)^{-9/2}\; G(\vka,\vkb,\vkc).
\end{equation}
The observationally relevant non-Gaussianity parameter~$\fnl$ is 
introduced through the relation
\begin{equation}
\cR(\eta, {\bm x})=\cR_{_{\mathrm{G}}}(\eta, {\bm x})
-\frac{3\,\fnl}{5}\, 
\l[\cR_{_{\mathrm{G}}}^2(\eta,{\bm x})
-\l\langle\cR_{_{\mathrm{G}}}^2(\eta, {\bm x})\r\rangle\r],
\label{eq:fnl-i}
\end{equation}
where $\cR_{_{\mathrm{G}}}$ denotes the Gaussian quantity, and the factor 
of $3/5$ arises due to the relation between the Bardeen potential and the
curvature perturbation during the matter dominated epoch.
Upon making use of the corresponding relation between $\cR$ and 
$\cR_{_{\mathrm{G}}}$ in Fourier space and the Wick's theorem, one obtains
that~\cite{maldacena-2003,ng-ncsf,ng-reviews}
\begin{eqnarray}
\langle \hat \cR_{\vka} \hat \cR_{\vkb} \hat \cR_{\vkc} \rangle
&=& -\frac{3\,\fnl}{10}\; (2\,\pi)^{4}\; (2\,\pi)^{-3/2}\;
\f{1}{k_{1}^3\, k_{2}^3\,k_{3}^3\,}\,
\delta^{(3)}(\vka+\vkb+\vkc)\nn\\
& &\times\l[k_1^{3}\; {\cal P}_{_{\rm S}}(k_2)\; {\cal P}_{_{\rm S}}(k_3)
+{\rm two~permutations}\r].
\end{eqnarray}
This expression can then be utilized to arrive at the following relation 
between the non-Gaussianity parameter $\fnl$ and the scalar bi-spectrum 
$\cB_{_{\rm S}}(\vka,\vkb,\vkc)$ or, equivalently, the quantity
$G(\vka,\vkb,\vkc)$ that we have 
introduced~\cite{martin-2012a,hazra-2012}:
\begin{eqnarray}
\fnl(\vka,\vkb,\vkc)
&=&-\frac{10}{3}\; (2\,\pi)^{-4}\;
\l(k_{1}\, k_{2}\,k_{3}\r)^3\;  G(\vka,\vkb,\vkc)\nn\\
& &\times\l[k_1^{3}\; {\cal P}_{_{\rm S}}(k_2)\; {\cal P}_{_{\rm S}}(k_3)
+{\rm two~permutations}\r]^{-1}.\label{eq:fnl-d}
\end{eqnarray}
In particular, in the equilateral limit, i.e. when $\ka=\kb=\kc=k$, this
expression simplifies to
\begin{equation}
\fnl^{\mathrm{eq}}(k)=-\frac{10}{9}\; \f{1}{(2\,\pi)^{4}}\; 
\f{k^6\; G(k)}{{{\cal P}_{_{\rm S}}^{2}(k)}}.
\label{eq:fnl-eq}
\end{equation}
Two points need to be emphasized here regarding the  
above expressions for the quantities $\fnl$ and $\fnl^{\mathrm{eq}}$.
Firstly, it ought to be noted that the non-Gaussianity $\fnl$ parameter 
introduced in Eq.~(\ref{eq:fnl-i}) [as well as the expression arrived at 
in Eq.~(\ref{eq:fnl-d})] corresponds to the so-called local form of the 
parameter.
Secondly, the quantity $\fnl^{\mathrm{eq}}$ corresponds to the equilateral 
limit of the {\it local}\/ $\fnl$ parameter, and is intrinsically different 
from a similar parameter introduced when the bi-spectrum has the equilateral 
form (in this context, see, for instance, Ref.~\cite{wmap-2012}).


\subsection{The scalar bi-spectrum in the Maldacena formalism}
 
In the Maldacena formalism~\cite{maldacena-2003}, the bi-spectrum 
is evaluated using the standard rules of perturbative quantum field 
theory, based on the interaction Hamiltonian that depends cubically 
on the curvature perturbation.
It can be shown that the bi-spectrum that results from the interaction 
Hamiltonian can be expressed 
as~\cite{maldacena-2003,ng-ncsf,ng-reviews,ng-f1,ng-f,martin-2012a,hazra-2012}
\begin{eqnarray}
G(\vka,\vkb,\vkc)
&\equiv & \sum_{C=1}^{7}\; G_{_{C}}(\vka,\vkb,\vkc)\nn\\
&\equiv & \Mp^2\; \sum_{C=1}^{6}\; 
\Biggl\{\l[f_{\vka}(\ee)\, f_{\vkb}(\ee)\,f_{\vkc}(\ee)\r]\; 
\cG_{_{C}}(\vka,\vkb,\vkc)\nn\\ 
& &+\l[f_{\vka}^{\ast}(\ee)\, f_{\vkb}^{\ast}(\ee)\,f_{\vkc}^{\ast}(\ee)\r]\;
\cG_{_{C}}^{\ast}(\vka,\vkb,\vkc)\Biggr\}
\nonumber \\ & &
+ G_{7}(\vka,\vkb,\vkc).\label{eq:G}
\end{eqnarray}
The quantities $\cG_{_{C}}(\vka,\vkb,\vkc)$ with $C =(1,6)$ correspond 
to the six terms in the interaction Hamiltonian, and are described by 
the integrals
\begin{eqnarray}
\cG_{1}(\vka,\vkb,\vkc)
&=&2\,i\,\int_{\ei}^{\ee} \d\eta\, a^2\, 
\epsilon_{1}^2\, \l(f_{\vka}^{\ast}\,f_{\vkb}'^{\ast}\,
f_{\vkc}'^{\ast}+{\rm two~permutations}\r),\label{eq:cG1}\\
\cG_{2}(\vka,\vkb,\vkc)
&=&-2\,i\;\l(\vka\cdot \vkb + {\rm two~permutations}\r)\,
\nonumber \\ & & \times
\int_{\ei}^{\ee} \d\eta\, a^2\, 
\epsilon_{1}^2\, f_{\vka}^{\ast}\,f_{\vkb}^{\ast}\,
f_{\vkc}^{\ast},\label{eq:cG2}\\
\cG_{3}(\vka,\vkb,\vkc)
&=&-2\,i\,\int_{\ei}^{\ee} \d\eta \, a^2\,
\epsilon_{1}^2\, \Biggl[\l(\f{\vka\cdot\vkb}{\kb^{2}}\r)\,
f_{\vka}^{\ast}\,f_{\vkb}'^{\ast}\, f_{\vkc}'^{\ast} \nonumber \\ & &
+ {\rm five~permutations}\Biggr],\label{eq:cG3}\\
\cG_{4}(\vka,\vkb,\vkc)
&=&i\,\int_{\ei}^{\ee} \d\eta\, a^2\,\epsilon_{1}\,
\epsilon_{2}'\, \l(f_{\vka}^{\ast}\,f_{\vkb}^{\ast}\,
f_{\vkc}'^{\ast}+{\rm two~permutations}\r),\label{eq:cG4}\\
\cG_{5}(\vka,\vkb,\vkc)
&=&\frac{i}{2}\,\int_{\ei}^{\ee} \d\eta\, 
a^2\, \epsilon_{1}^{3}\, \Biggl[\l(\f{\vka\cdot\vkb}{\kb^{2}}\r)\,
f_{\vka}^{\ast}\,f_{\vkb}'^{\ast}\, f_{\vkc}'^{\ast}
\nonumber \\ & & 
+ {\rm five~permutations}\Biggr],\label{eq:cG5}\\
\cG_{6}(\vka,\vkb,\vkc) 
&=&\frac{i}{2}\,\int_{\ei}^{\ee} \d\eta\, a^2\, 
\epsilon_{1}^{3}\,
\Biggl\{\l[\f{\ka^{2}\,\l(\vkb\cdot\vkc\r)}{\kb^{2}\,\kc^{2}}\r]\, 
f_{\vka}^{\ast}\, f_{\vkb}'^{\ast}\, f_{\vkc}'^{\ast}
\nonumber \\ & &
+ {\rm two~permutations}\Biggr\},\label{eq:cG6}
\end{eqnarray}
where $\epsilon_2$ is the second slow roll parameter that is defined
with respect to the first as follows: $\epsilon_2
=\d\, {\rm ln}\,\epsilon_1/\d N$.
The lower limit of the above integrals, viz. $\ei$, denotes a sufficiently
early time when the initial conditions are imposed on the modes. 
The additional, seventh term $G_{7}(\vka,\vkb,\vkc)$ arises due 
to the field redefinition, and its contribution to $G(\vka,\vkb,\vkc)$ 
is given by
\begin{equation}
G_{7}(\vka,\vkb,\vkc)
=\frac{\epsilon_{2}(\eta_{\rm e})}{2}\,
\l(\vert f_{\vkb}(\eta_{\rm e})\vert^{2}\, 
\vert f_{\vkc}(\eta_{\rm e})\vert^{2} 
+ {\rm two~permutations}\r).\label{eq:G7}
\end{equation} 


\section{The numerical computation of the scalar bi-spectrum}\label{sec:nc-sbs} 

In this section, after illustrating that the super-Hubble contributions
to the complete bi-spectrum during inflation proves to be negligible, we 
shall outline the methods that BINGO adopts to numerically evolve the 
equations governing the background and the perturbations, and eventually 
evaluate the inflationary scalar power and bi-spectra.
Also, we shall illustrate the extent of accuracy of BINGO by comparing the
results from the code with the expected form of the bi-spectrum in the 
equilateral limit in power law inflation and the analytical results that 
are available in the case of the Starobinsky 
model~\cite{martin-2012a,arroja-2011-2012}.
We shall also contrast the numerical results with the analytical results 
obtained under the slow roll approximation in the case of the popular 
quadratic potential for an arbitrary triangular configuration of the
wavevectors.


\subsection{The contributions to the bi-spectrum on super-Hubble 
scales}\label{sec:shc-bs}

It is clear from the expressions in the previous section that the evaluation 
of the bi-spectrum involves integrals over the mode $f_\vk$ and its derivative 
$f_\vk'$ as well as the slow roll parameters $\epsilon_1$, $\epsilon_2$ 
and the derivative $\epsilon_2'$.
While evaluating the power spectra, it is well known that, in single field
inflationary models, it suffices to evolve the curvature perturbation from 
an initial time when the modes are sufficiently inside the Hubble radius 
to a suitably late time when the amplitude of the curvature perturbation 
settles down to a constant value on super-Hubble scales (see, for example, 
Refs.~\cite{ne-ps}).
We shall illustrate that many of the contributions to the bi-spectrum prove 
to be negligible when the modes evolve on super-Hubble scales.
Interestingly, we shall also show that, those contributions to the 
bi-spectrum which turn out to be significant at late times when the 
modes are well outside the Hubble radius are canceled by certain other 
contributions that arise.
As a consequence, we shall argue that, numerically, it suffices to evaluate 
the integrals over the period of time during which the curvature perturbations 
have been conventionally evolved to arrive at the power spectra, viz. from
the sub-Hubble to the super-Hubble scales.
In fact, we had recently shown that, for cosmological modes that are on 
super-Hubble scales during preheating, the contributions to the bi-spectrum 
and, equivalently, to the non-Gaussianity parameter $\fnl$, prove to be 
negligible~\cite{hazra-2012}. 
The calculations we had presented in the work can be easily extended to 
the case of the modes on super-Hubble scales during inflation.
Therefore, we shall rapidly sketch the essentials arguments below, referring 
the readers to our recent work for further details~\cite{hazra-2012}.


\subsubsection{Evolution of $f_\vk$ on super-Hubble scales}

During inflation, when the modes are on super-Hubble scales, it is well 
known that the solution to $f_\vk$ can be written as~\cite{texts,reviews}
\begin{equation}
f_\vk\simeq A_\vk+ B_\vk\, \int ^{\eta}
\f{\d{\tilde \eta}}{z^2({\tilde \eta})},
\end{equation}
where $A_\vk$ and $B_\vk$ are $\vk$-dependent constants which are determined 
by the initial conditions imposed on the modes in the sub-Hubble limit.
The first term involving $A_\vk$ is the growing mode, which is actually a 
constant, while the term containing $B_\vk$ represents the decaying mode. 
Therefore, on super-Hubble scales, the mode $f_\vk$ simplifies to 
\begin{equation}
f_\vk\simeq A_\vk.\label{eq:fk-shs}
\end{equation}
Moreover, the leading non-zero contribution to its derivative is determined
by the decaying mode, and is given by
\begin{equation}
f_\vk'\simeq \f{B_\vk}{z^2}
=\f{{\bar B}_\vk}{a^2\, \epsilon_1},\label{eq:fkp-shs}
\end{equation}
where we have set ${\bar B}_\vk=B_\vk/(2\, \Mp^2)$.

\par 

It is now a matter of making use of the above solutions for $f_\vk$ and 
$f_\vk'$ to determine the super-Hubble contributions to the bi-spectrum 
during inflation.


\subsubsection{The various contributions}

\begin{figure}[!t]
\begin{center}
\resizebox{420pt}{280pt}{\includegraphics{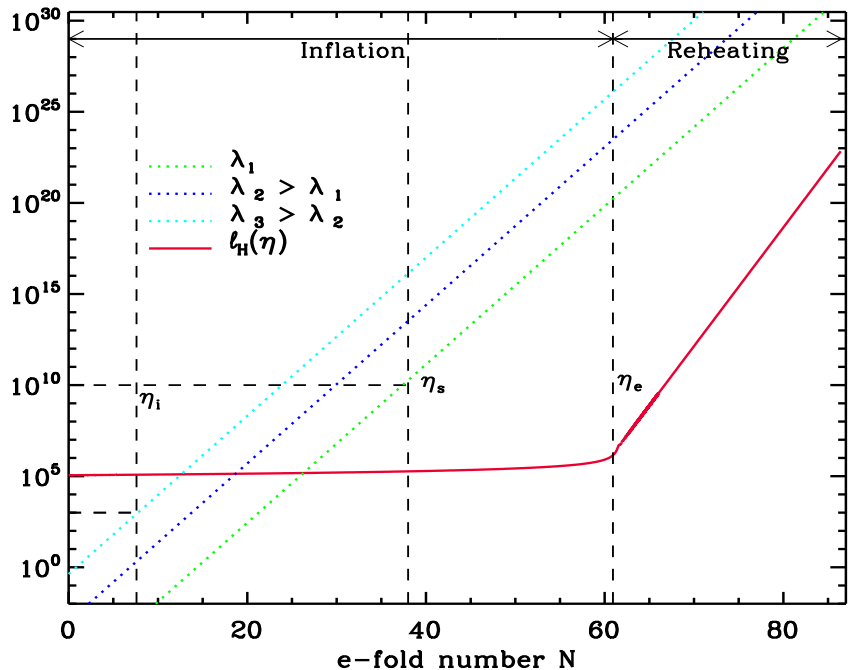}}
\end{center}
\vskip -10pt
\caption{\label{fig:eta}The evolution of three wavelengths, say, 
$\lambda_1< \lambda_2<\lambda_3$ (corresponding to the wavenumbers 
$k_1 > k_2 > k_3$) and the Hubble radius $\ell_{_{\rm H}}=H^{-1}$ 
have been plotted as a function of e-folds $N$ (in green, blue, 
cyan and red, respectively) during the epochs of inflation and reheating 
for the case of the conventional quadratic potential.
Note that $\ei$ corresponds to the time when the largest of the three 
wavelengths, viz. $\lambda_3$ in this case, satisfies the condition 
$k_3/(a\, H)\simeq 100$, while $\es$ denotes the time when the smallest
of the three wavelengths, i.e. $\lambda_1$, is sufficiently outside 
the Hubble radius such that, say, $k_1/(a\, H)\simeq 10^{-5}$.
Inflation terminates at the time $\ee$, when the epoch of reheating 
is expected to begin.}
\end{figure}

To begin with, note that, each of the integrals 
${\cal G}_{_{C}}(\vka,\vkb,\vkc)$, where $C=(1,6)$, can be divided
into two parts as follows:
\begin{eqnarray}
\cG_{_{C}}(\vka,\vkb,\vkc)
=\cG_{_{C}}^{\rm is}(\vka,\vkb,\vkc)
+\cG_{_{C}}^{\rm se}(\vka,\vkb,\vkc).
\end{eqnarray}
The integrals in the first term $\cG_{_{C}}^{\rm is}(\vka,\vkb,\vkc)$ run 
from the earliest time (i.e. $\ei$) when the smallest of the three 
wavenumbers $k_1$, $k_2$ and $k_3$ is sufficiently inside the Hubble radius 
[typically corresponding to $k/(a\, H)\simeq 100$] to the time (say, $\es$) 
when the largest of the three wavenumbers is well outside the Hubble radius 
[say, when $k/(a\, H)\simeq 10^{-5}$].
Then, evidently, the second term $\cG_{_{C}}^{\rm se}(\vka,\vkb,\vkc)$ will 
involve integrals which run from the latter time $\es$ to the end of inflation 
at~$\ee$ (in this context, see Fig.~\ref{fig:eta}).
In what follows, we shall discuss the various contributions to the bi-spectrum
due to the terms $\cG_{_{C}}^{\rm se}(\vka,\vkb,\vkc)$.
We shall show that the corresponding contribution either remains small or, when
it proves to be large, it is exactly canceled by another contribution to the
bi-spectrum.

\par 

Let us first focus on the fourth term $G_{4}(\vka,\vkb,\vkc)$ since it 
has often been found to lead to the largest contribution to the bi-spectrum 
when deviations from slow roll occur~\cite{ng-f1,ng-f,martin-2012a,hazra-2012}. 
As the slow roll parameters turn large towards the end of inflation, we can 
expect this term to contribute significantly at late times. 
However, as we shall quickly illustrate, such a late time contribution is 
exactly canceled by the contribution from $G_{7}(\vka,\vkb,\vkc)$ which 
arises due to the field redefinition~\cite{hazra-2012}.
Upon using the form~(\ref{eq:fk-shs}) of the mode $f_{\vk}$ and 
its derivative~(\ref{eq:fkp-shs}) on super-Hubble scales in 
the expression~(\ref{eq:cG4}), the integral involved can be 
trivially carried out with the result that the corresponding 
contribution to the bi-spectrum can be expressed as
\begin{eqnarray}
G_{4}^{\rm se}(\vka,\vkb,\vkc)
&\simeq& i\, \Mp^2\, \l[\epsilon_2(\ee)-\epsilon_2(\es)\r]\,
\biggl[\vert A_{\vka}\vert^2\, \vert A_{\vkb}\vert^2\,
\l(A_{\vkc}\, {\bar B}_{\vkc}^{\ast}-A_{\vkc}^{\ast}\, 
{\bar B}_{\vkc}\r)\nn\\
& &+\,{\rm two~permutations}\biggr].\label{eq:G4-d-prh}
\end{eqnarray}
The Wronskian corresponding to the equation of
motion~(\ref{eq:de-fk}) and the standard Bunch-Davies initial
condition leads to the relation $(A_\vk\, {\bar B}_\vk^{\ast}\
-A_\vk^{\ast}\, {\bar B}_\vk)=i/(2\, \Mp^2)$, which can then be
utilized to arrive at the following simpler
expression~\cite{hazra-2012}:
\begin{eqnarray}
G_{4}^{\rm se}(\vka,\vkb,\vkc)
&\simeq& -\f{1}{2}\, \l[\epsilon_2(\ee)-\epsilon_2(\es)\r]\nn\\
& &\times\,
\biggl[\vert A_{\vka}\vert^2\, \vert A_{\vkb}\vert^2
+\,~{\rm two~permutations}\biggr].\label{eq:G4-at-es}
\end{eqnarray}
The first of these terms involving the value of $\epsilon_2$ at the
end of inflation {\it exactly}\/ cancels the contribution 
$G_{7}(\vka,\vkb,\vkc)$ [with $f_\vk$ set to $A_\vk$ in Eq.~(\ref{eq:G7})] 
that arises due to the field redefinition. 
But, the remaining contribution cannot be ignored and needs to be 
taken into account.
It is useful to note that this term is essentially the same as the 
one due to the field redefinition, but which is now evaluated on 
super-Hubble scales (i.e. at $\es$) rather than at the end of inflation.
In other words, if we consider the fourth and the seventh terms 
together, it is equivalent to evaluating the contribution to the 
bi-spectrum corresponding to $\cG_{4}^{\rm is}(\vka,\vkb,\vkc)$, 
and adding to it the contribution due to the seventh term
$G_{7}(\vka,\vkb,\vkc)$ evaluated at $\es$ rather than at the end of
inflation.

\par

Let us now turn to the contribution due to the second term, which 
can occasionally prove to be comparable to the contribution due 
to the fourth term~\cite{martin-2012a}.
Upon making use of the behavior of the mode $f_\vk$ and its derivative 
on super-Hubble scales in the integral~(\ref{eq:cG2}), one obtains 
the contribution to the bi-spectrum due to 
$\cG_{2}^{\rm se}(\vka,\vkb,\vkc)$ to be~\cite{martin-2012a,hazra-2012}
\begin{eqnarray}
G_2^{\rm se}(\vka,\vkb,\vkc)
\!&=&\!-2\, i\,\Mp^2\,\l(\vka\cdot \vkb + {\rm two~permutations}\r)\nn\\
& &\times\, \vert A_{\vka}\vert^2\, 
\vert A_{\vkb}\vert^2\, 
\vert A_{\vkc}\vert^2\,\l[I_2(\ee,\es)-I_2^{\ast}(\ee,\es)\r],
\end{eqnarray}
where the quantity $I_2(\ee,\es)$ is described by the integral
\begin{equation}
I_2(\ee,\es)=\int_{\es}^{\ee} \d\eta\; a^2\, \epsilon_{1}^2.
\end{equation}
Note that, due to the quadratic dependence on the scale factor, actually,
$I_2(\ee,\es)$ is a rapidly growing quantity at late times.
However, the complete super-Hubble contribution to the bi-spectrum
vanishes identically since the integral $I_2(\ee,\es)$ is a purely
real quantity~\cite{hazra-2012}.
Hence, in the case of the second term, it is sufficient to evaluate the 
contribution to the bi-spectrum due to $\cG_{2}^{\rm is}(\vka,\vkb,\vkc)$.

\par 

Due to their structure, one finds that the first and the third terms
and the fifth and the sixth terms can be evaluated together. 
On super-Hubble scales, one can easily show that the contributions due 
to the first and the third terms can be written as
\begin{eqnarray}
\label{eq:G13-shc}
G_1^{\rm es}(\vka,\vkb,\vkc) + G_3^{\rm es}(\vka,\vkb,\vkc)
&=&2\, i\,\Mp^2\,\Biggl[\l(1-\f{\vka\cdot\vkb}{\kb^{2}}
-\f{\vka\cdot\vkc}{\kc^{2}}\r)\,\vert A_{\vka}\vert^2\nn\\
& &\times\l(A_{\vkb}\, {\bar B}_{\vkb}^{\ast}\, A_{k_{3}}
{\bar B}_{\vkc}^{\ast}-A_{\vkb}^{\ast} {\bar B}_{\vkb}\,
A_{\vkc}^{\ast}\, {\bar B}_{\vkc}\r)\nn\\
& &+~{\rm two~permutations}\Biggr]\;I_{13}(\ee,\es),
\end{eqnarray}
while the corresponding contributions due to the fifth and the sixth 
terms can be obtained to be
\begin{eqnarray}
G_5^{\rm se}(\vka,\vkb,\vkc)+G_6^{\rm se}(\vka,\vkb,\vkc)
&=&\f{i\,\Mp^2}{2}\nn\\
& &\times\biggl[\l(\f{\vka\cdot\vkb}{\kb^{2}}
+\f{\vka\cdot\vkc}{\kc^{2}}
+\f{\ka^{2}\,\l(\vkb\cdot\vkc\r)}{\kb^{2}\,\kc^{2}}\r)\nn\\
& &\times\,\vert A_{\vka}\vert^2\nn\\
& &\times \l(A_{\vkb} {\bar B}_{\vkb}^{\ast}\, A_{\vkc}
{\bar B}_{\vkc}^{\ast}-A_{\vkb}^{\ast} {\bar B}_{\vkb}\,
A_{\vkc}^{\ast}\, {\bar B}_{\vkc}\r)\nn\\
& &+~{\rm two~permutations}\biggr]\;I_{56}(\ee,\es),
\label{eq:G56-shc}
\end{eqnarray}
where the quantities $I_{13}(\ee,\es)$ and $I_{56}(\ee,\es)$ are described
by the integrals
\begin{equation}
I_{13}(\ee,\es)=\int_{\es}^{\ee} \f{\d\eta}{a^2}
\end{equation}
and 
\begin{equation}
I_{56}(\ee,\es)=\int_{\es}^{\ee} \f{\d\eta}{a^2}\, \epsilon_1.
\end{equation}
Hence, the non-zero, super-Hubble contribution to the bi-spectrum is 
determined by the complete contribution due to the first, the third, 
the fifth and the sixth terms arrived at above.
In order to illustrate that this contribution is insignificant, we shall 
now turn to estimating the amplitude of the corresponding contribution 
to the non-Gaussianity parameter~$\fnl$.


\subsection{An estimate of the super-Hubble contribution to the 
non-Gaussianity parameter}
 
Let us restrict ourselves to the equilateral limit for simplicity.
In such a case, the super-Hubble contributions due to the first, the 
third, the fifth and the sixth terms to the non-Gaussianity 
parameter~$\fnl$ can be obtained by substituting the 
quantities~(\ref{eq:G13-shc}) and~(\ref{eq:G56-shc}) above in the
expression~(\ref{eq:fnl-eq}).  
It is straightforward to show that the corresponding $\fnl$ is given by 
\begin{eqnarray}
\label{eq:fnl-shs}
\fnl^{\rm eq\, (se)}(k)
&\simeq& -\f{5\, i\,\Mp^2}{18}\, 
\l(\f{A_{\bm k}^2\, {\bar B}_{\bm k}^{\ast}{}^2\,
- A_{\bm k}^{\ast}{}^2\, {\bar B}_{\bm k}^2}{\vert A_{\bm k}\vert^2}\r)\nn\\
& &\times\,\l[12\, I_{13}(\ee,\es)-\f{9}{4}\, I_{56}(\ee,\es)\r],
\end{eqnarray}
where we have made use of the fact that $f_\vk\simeq A_\vk$ at late 
times in order to arrive at the power spectrum.

\par 

To estimate the above super-Hubble contribution to the non-Gaussianity 
parameter $\fnl^{\rm eq}$, let us focus on inflation of the power law 
form.
In power law inflation, the scale factor is given by 
\begin{equation}
a(\eta)=a_1\, \l(\f{\eta}{\eta_1}\r)^{\gamma+1},\label{eq:pli}
\end{equation}
with $a_1$ and $\eta_1$ being constants, while $\gamma$ is a free index. 
It is useful to note that, in such a case, the first slow roll parameter is 
a constant, and is given by $\epsilon_1=(\gamma +2)/(\gamma +1)$. 
Under these conditions, the quantity within the brackets involving $A_\vk$
and ${\bar B}_\vk$ in the expression above can be easily evaluated (in this
context, see Ref.~\cite{hazra-2012}) to arrive at
\begin{eqnarray}
\fnl^{\rm eq\, (se)}(k)
&=& \frac{5}{72\,\pi}\; \l[12-\f{9\, (\gamma+2)}{\gamma+1}\r]\,
\Gamma^2\l(\gamma +\f{1}{2}\r)\,
2^{2\,\gamma+1}\, \l(2\,\gamma+1\r)\, (\gamma+2)\nn\\ 
& &\times\,\l(\gamma+1\r)^{-2\,(\gamma+1)}\,
\sin\,(2\,\pi\,\gamma)\,
\l[1-\f{H_{\rm s}}{H_{\rm e}}\,{\rm e}^{-3\,(\Ne-\Ns)}\r]\nn\\
& &\times\,\l(\frac{k}{a_{\rm s}\, H_{\rm s}}\r)^{-(2\,\gamma+1)}.
\end{eqnarray}
It should be mentioned that, in arriving at this expression, for convenience, 
we have set $\eta_1$ to be $\eta_s$, which corresponds to $a_1$ being 
$a_{\rm s}$, viz. the scale factor at $\es$.
Moreover, while $\Ns$ and $\Ne$ denote the e-folds corresponding to $\es$ and 
$\ee$, $H_{\mathrm{s}}$ and $H_{\mathrm{e}}$ represent the Hubble scales at 
these times, respectively.
Recall that, $\es$ denotes the conformal time when the largest wavenumber 
of interest, say, $k_{\mathrm{s}}$, is well outside the Hubble radius, i.e. 
when $k_{\mathrm{s}}/(a\, H)\simeq 10^{-5}$.
Since $(\Ne-\Ns)$ is expected to be at least $40$ for the smallest 
cosmological scale, it is clear that the factor involving $\exp-[3\,
(\Ne-\Ns)]$ can be completely neglected.
Observations point to the fact that $\gamma\lesssim-2$. 
Therefore, if we further assume that $\gamma=-(2+\varepsilon)$, where 
$\varepsilon \simeq 10^{-2}$, we find that the above estimate for the 
non-Gaussianity parameter reduces to
\begin{equation}
\fnl^{\rm eq\,(se)}(k) \lesssim -\f{5\,\varepsilon^2}{9}\,
\l(\frac{k_{\rm s}}{a_{\rm s}\, H_{\rm s}}\r)^{3}\simeq -10^{-19},
\label{eq:fnl-shs-ub}
\end{equation}
where, in obtaining the final value, we have set $k_{\mathrm{s}}/(a_{\mathrm{s}}\, 
H_{\mathrm{s}}) = 10^{-5}$.
The inequality above arises due to the fact that, for larger scales, i.e. 
when $k<k_{\mathrm{s}}$, $k/(a\, H)<10^{-5}$ at $\es$.  
In models involving the canonical scalar field, the smallest values of $\fnl$ 
are typically generated in slow roll inflationary scenarios, wherein the 
non-Gaussianity parameter has been calculated to be of the order of the first 
slow roll parameter~\cite{maldacena-2003,ng-reviews}.  
The above estimate clearly points to fact that the super-Hubble contributions 
to the complete bi-spectrum and the non-Gaussianity parameter $\fnl$ can be 
entirely ignored.

\par 

In summary, to determine the scalar bi-spectrum, it suffices to evaluate the
contributions to the bi-spectrum due to the quantities $\cG_{_{C}}^{\rm is}
(\vka,\vkb,\vkc)$, with $C=(1,6)$, which involve integrals running from the
initial time $\ei$ to the time $\es$ when the smallest of the three modes 
reaches super-Hubble scales (cf. Fig.~\ref{fig:eta}).
Further, the addition of the contribution due to the field redefinition 
evaluated at $\es$ ensures that no non-trivial super-Hubble contributions
are ignored. 
In the following sub-section, with the help of a specific example, we shall 
also corroborate these conclusions numerically.


\subsection{Details of the numerical methods}\label{subsec:dnm}

The scalar bi-spectrum and the parameter $\fnl$ can be easily evaluated
analytically in the slow roll inflationary scenario~\cite{maldacena-2003}.
However, barring some exceptional 
cases~\cite{martin-2012a,arroja-2011-2012,hu-2010-2011}, 
it often proves to be difficult to evaluate the bi-spectrum analytically when 
departures from slow roll occur.
Hence, one has to resort to numerical computations in such cases.
  
\par

BINGO solves the background as well as the perturbation equations using 
RKSUITE~\cite{rks}, which is a publicly available routine to solve ordinary 
differential equations.
We shall treat the number of e-folds as the independent variable, which 
allows for an efficient and accurate computation.
To obtain the power spectrum, we impose the standard Bunch-Davies initial 
conditions on the perturbations when the modes are well inside the Hubble 
radius, and evolve them until suitably late times. 
Typically, in the case of smooth inflaton potentials, it suffices to evolve 
the modes $f_\vk$ from an initial time when $k/(a\, H) = 100$.
However, in certain models wherein the potentials contain oscillatory terms, 
the modes may have to be evolved from deeper inside the Hubble radius.
For example, in the case of the axion monodromy model that we shall discuss
in due course (see Sub-sec.~\ref{subsec:oip}), for values of the model 
parameters of our interest, we find that we need to evolve the modes from an 
initial time when $k/(a\, H) \simeq 250$, so that the resonance that occurs 
in the model due to the oscillations in the potential is 
captured~\cite{flauger-2010-2011,aich-2011,ng-f1,ng-f}. 
The scalar power spectra that arise in the various models of our interest, as 
displayed in Fig.~\ref{fig:sps-all}, have all been evaluated at super-Hubble 
scales, say, when $k/(a\,H) \simeq 10^{-5}$, which is typically when the 
amplitude of the curvature perturbations freeze in.

\par 

Having obtained the behavior of the background and the modes, BINGO carries 
out the integrals involved in arriving at the bi-spectrum using the method 
of adaptive quadrature~\cite{gq}. 
It is useful to note that, in the equilateral limit of the bi-spectrum,
which we shall largely concentrate on in this work, we can evolve each of 
the modes of interest independently and calculate the integrals for the 
modes separately.
The integrals $\cG_{n}$ actually contain a cut off in the sub-Hubble
limit, which is essential for singling out the perturbative 
vacuum~\cite{maldacena-2003,ng-ncsf,ng-reviews}. 
Numerically, the presence of the cut off is fortunate since it controls 
the contributions due to the continuing oscillations that would otherwise 
occur. 
We should highlight here that such a cut off procedure was 
originally introduced in the earliest numerical efforts to evaluate the 
scalar bi-spectrum in situations involving deviations from slow roll (in this 
context, see Refs.~\cite{ng-f1}).
Generalizing the cut off that is often introduced analytically in the slow 
roll case, we shall impose a cut off of the form $\exp-[\kappa\; k/(a\, H)]$, 
where $\kappa$ is a small parameter. 
In the previous two sub-sections, we had discussed as to how the integrals 
need to be carried out from the early time $\ei$ when the largest scale is 
well inside the Hubble radius to the late time $\es$ when the smallest scale 
is sufficiently outside. 
If one is focusing on the equilateral configuration, rather than integrate 
from $\ei$ to $\es$, it suffices to compute the integrals for the modes from 
the time when each of them satisfy the sub-Hubble condition, say, $k/(a\,H)= 
100$, to the time when they are well outside the Hubble radius, say, when 
$k/(a\,H) = 10^{-5}$.
In other words, one carries out the integrals exactly over the period the 
modes are evolved to obtain the power spectrum.
The presence of the cut-off ensures that the contributions at early times, 
i.e. near $\ei$, are negligible.
Furthermore, it should be noted that, in such a case, the corresponding 
super-Hubble contribution to $\fnl^{\rm eq}$ will saturate the 
bound~(\ref{eq:fnl-shs-ub}) in power law inflation for all the modes.

\par

\begin{figure}[!t]
\begin{center}
\psfrag{G-all}[0][1][1.0]{$k^6\, \vert G_{n}(k)\vert$} 
\psfrag{delta}[0][1][1.0]{$\kappa$}
\resizebox{420pt}{280pt}{\includegraphics{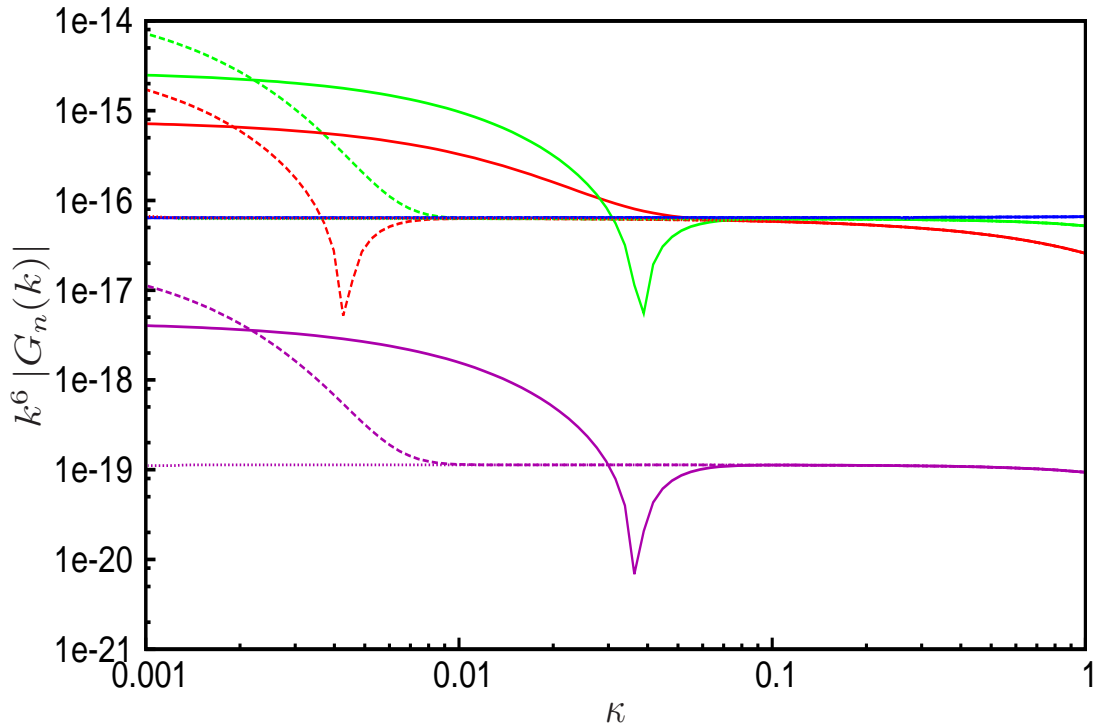}}
\end{center}
\vskip -10pt
\caption{\label{fig:G-delta} The quantities $k^6$ times 
the absolute values of $G_1+G_3$ (in green), $G_2$ (in red), $G_4+G_7$ 
(in blue) and $G_5+G_6$ (in purple) have been plotted as a function of 
the cut off parameter $\kappa$ for a given mode in the case of the 
conventional, quadratic inflationary potential.
Note that these values have been arrived at with a fixed upper limit 
[viz. corresponding to $k/(a\, H)=10^{-5}$] for the integrals involved.
The solid, dashed and the dotted lines correspond to integrating from 
$k/(a\,H)$ of $10^2$, $10^3$ and $10^4$, respectively.
It is clear that the results converge for $\kappa=0.1$, which suggests it
to be an optimal value.
While evaluating the bi-spectrum for the other models, we shall choose to
work with a $\kappa$ of $0.1$, and impose the initial conditions as well as 
carry out the integrals from $k/(a\, H)$ of $10^2$ (barring the case of the 
axion monodromy model, as we have discussed in the text).
An additional point that is worth noticing is the fact the term $G_4+G_7$
seems to be hardly dependent of the cut-off parameter~$\kappa$.
This can possibly be attributed to the dependence of $G_4$ on $\epsilon_2'$,
which can be rather small during slow roll, thereby effectively acting as a
cut off.
Due to this reason, hereafter, we shall not introduce the cut-off while 
calculating $G_4$.}
\end{figure}

\begin{figure}[!t]
\begin{center}
\psfrag{G}[0][1][1.0]{$k^6\, \vert G_{n}(k)\vert$} 
\psfrag{Nshs}[0][1][0.9]{$N_{_{\mathrm{S}}}$}
\psfrag{kah}[0][1][0.9]{$k/(a\,H)$}
\resizebox{420pt}{280pt}{\includegraphics{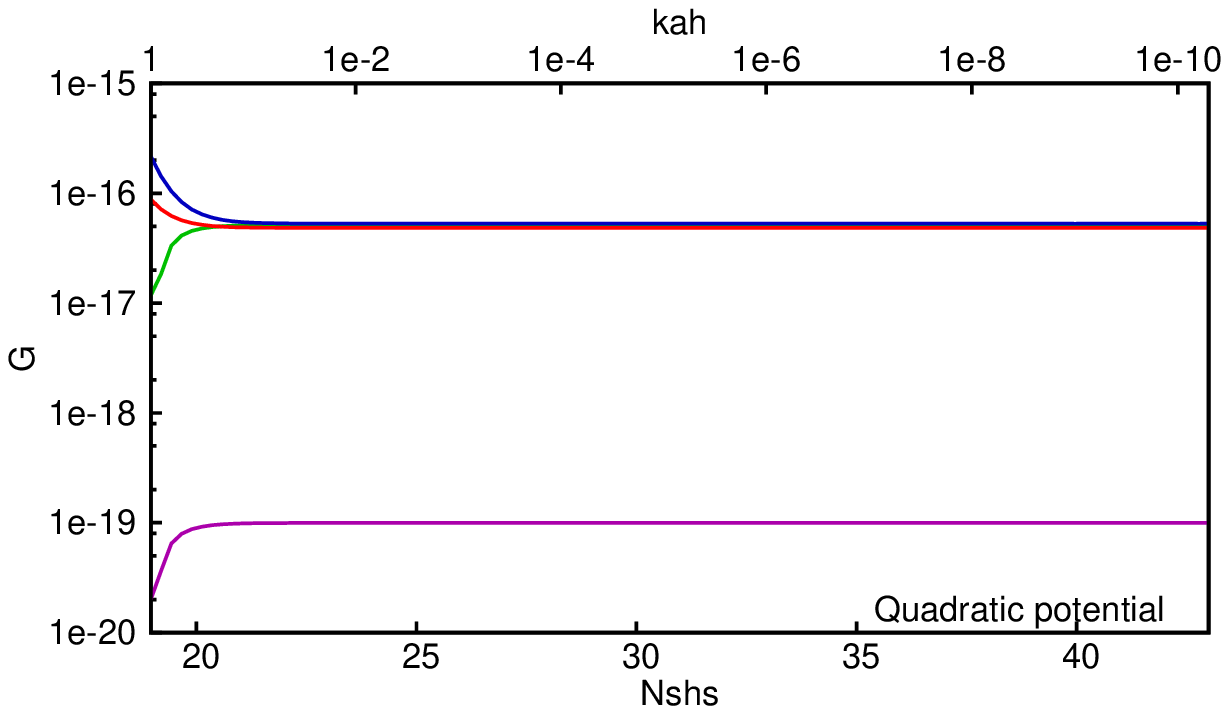}}
\end{center}
\vskip -10pt
\caption{\label{fig:G-Nshs-all} The quantities $k^6$ times 
the absolute values of $G_1+G_3$, $G_2$, $G_4+G_7$ and $G_5+G_6$ have been 
plotted (with the same choice of colors as in the previous figure) as a 
function of the upper limit of the integrals involved for a given mode in 
the case of the quadratic potential.
Evidently, the integrals converge fairly rapidly to their final values
once the mode leaves the Hubble radius.
The independence of the results on the upper limit support the conclusions
that we had earlier arrived at analytically in the last sub-section, 
viz. that the super-Hubble contributions to the bi-spectrum are entirely
negligible.
Another point that we should stress here is the fact that, since the 
quadratic potential only admits slow inflation, it is the $G_7$ term that 
dominates in the combination $G_4+G_7$.}
\end{figure}

With the help of a specific example, let us now illustrate that, for a 
judicious choice of~$\kappa$, the results from BINGO are largely 
independent of the upper and the lower limits of the integrals.
In fact, working in the equilateral limit, we shall demonstrate these 
points in two steps for the case of the standard quadratic 
potential [see Eq.~(\ref{eq:qp})]. 
Firstly, focusing on a specific mode, we shall fix the upper limit of 
the integral to be the time when $k/(a\,H)= 10^{-5}$.
Evolving the mode from different initial times, we shall evaluate the 
integrals involved from these initial times to the fixed final time for 
different values of $\kappa$.
This exercise helps us to identify an optimal value for $\kappa$ when 
we shall eventually carry out the integrals from $k/(a\,H)= 100$.
Secondly, upon choosing the optimal value for $\kappa$ and integrating 
from $k/(a\,H)= 100$, we shall calculate the integrals for different 
upper limits.
For reasons outlined in the previous two sub-sections, it proves to be
necessary to consider the contributions to the bi-spectrum due to 
the fourth and the seventh terms together.
Moreover, since the first and the third, and the fifth and the sixth,
have similar structure in the equilateral limit, it turns out to be 
convenient to club these terms as we have discussed before. 
In Fig.~\ref{fig:G-delta}, we have plotted the value of $k^6$ times 
the different contributions to the bi-spectrum, viz. $G_{1}+G_{3}$, 
$G_{2}$, $G_4+G_7$ and $G_5+G_6$, as a function of $\kappa$ when the 
integrals have been carried out from $k/(a\,H)$ of $10^2$, $10^3$ and 
$10^4$ for a mode which leaves the Hubble radius around $53$ e-folds 
before the end of inflation in the case of the quadratic potential.
The figure suggests that the term $G_4+G_7$ is fairly independent of 
the cut-off parameter $\kappa$.
This can be attributed to the fact that the integral $\cG_4$ depends 
on the quantity $\epsilon_2'$, which effectively behaves as a cut off. 
For this reason, actually, we shall {\it not}\/ introduce the cut off 
while evaluating the term $G_4$.
Moreover, the figure clearly indicates $\kappa=0.1$ to be a highly 
suitable value for the other terms.
A larger $\kappa$ leads to a sharper cut off reducing the value of the 
integrals.
One could work with a smaller $\kappa$, in which case, the figure suggests
that, one would also need to necessarily integrate from deeper inside the 
Hubble radius.
In Fig.~\ref{fig:G-Nshs-all}, after fixing $\kappa$ to be $0.1$ and, with
the initial conditions imposed at $k/(a\, H)=10^2$, we have plotted the 
four contributions to the bi-spectrum for a mode that leaves the Hubble
radius at $50$ e-folds before the end of inflation as a function of the 
upper limit of the integrals. 
It is evident from the figure that the values of the integrals converge 
quickly once the mode leave the Hubble radius.
For efficient numerical integration, as in the case of the power spectrum,
we have chosen the super-Hubble limit to correspond to $k/(a\,H)=10^{-5}$.
We have repeated similar tests for the other models of our interest too.
These tests confirm the conclusions that we have arrived at above,
indicating the robustness of the numerical methods and procedures
that we have adopted.
It is important that we stress three points further at this stage of our 
discussion.
Firstly, in situations involving departures from slow roll, the dominant  
contribution, viz. the $G_4+G_7$ term, does not require the introduction
of the cut off at all.
Secondly, in the absence of analytical expressions to compare the numerical
results with, it is imperative that a suitable value for the cut off parameter
$\kappa$ is arrived at by repeating the exercise that we have carried out in 
Fig.~\ref{fig:G-delta} for the specific model of interest.
In this context, it is further important to appreciate the fact that the 
integrals involved will depend on both the cut off parameter $\kappa$ as 
well as the initial value of $k/(a\, H)$. 
A complete check seems to  be mandatory in order to ensure that a unique 
result is arrived at.
(As we mentioned above, we have indeed carried out this exercise for the 
models that we consider here.)
Lastly, we also need to emphasize that the above comments 
and conclusions would not apply to certain special cases wherein the 
amplitude of the curvature perturbation can evolve on super-Hubble 
scales such as, say, in ultra slow roll inflation (in this context, 
see, for instance, Ref.~\cite{martin-2012b}).

\par

Before we go on to consider the bi-spectra generated in the inflationary
models of our interest, in order to demonstrate the accuracy of BINGO, we 
shall compare the numerical results from the code with the analytical results 
that can be arrived at in three situations. 
In the following sub-sections, we shall compare the results from BINGO 
with: (i)~the spectral dependence that can be arrived at in the context 
of power law inflation in the equilateral limit, (ii)~the slow roll 
results as applied to the case of the conventional quadratic potential 
for a generic triangular configuration of the wavevectors, and (iii)~the
analytic results that can be obtained for the Starobinsky model in the
equilateral limit.


\subsection{Comparison with the analytical results in the case of
power law inflation} 

The first case that we shall discuss is power law inflation wherein,
as we shall soon outline, the spectral shape of the non-zero contributions 
to the bi-spectrum can be easily arrived at in the equilateral limit.

\par

\begin{figure}[!t]
\begin{center}
\psfrag{k6g}[0][1][1.5]{$k^6\, \vert G_{n}(k)\vert$} 
\psfrag{kmpc}[0][1][1.5]{$k$}
\psfrag{fnl}[0][1][1.5]{$\fnl^{\mathrm{eq}}$}
\resizebox{460pt}{310pt}{\includegraphics{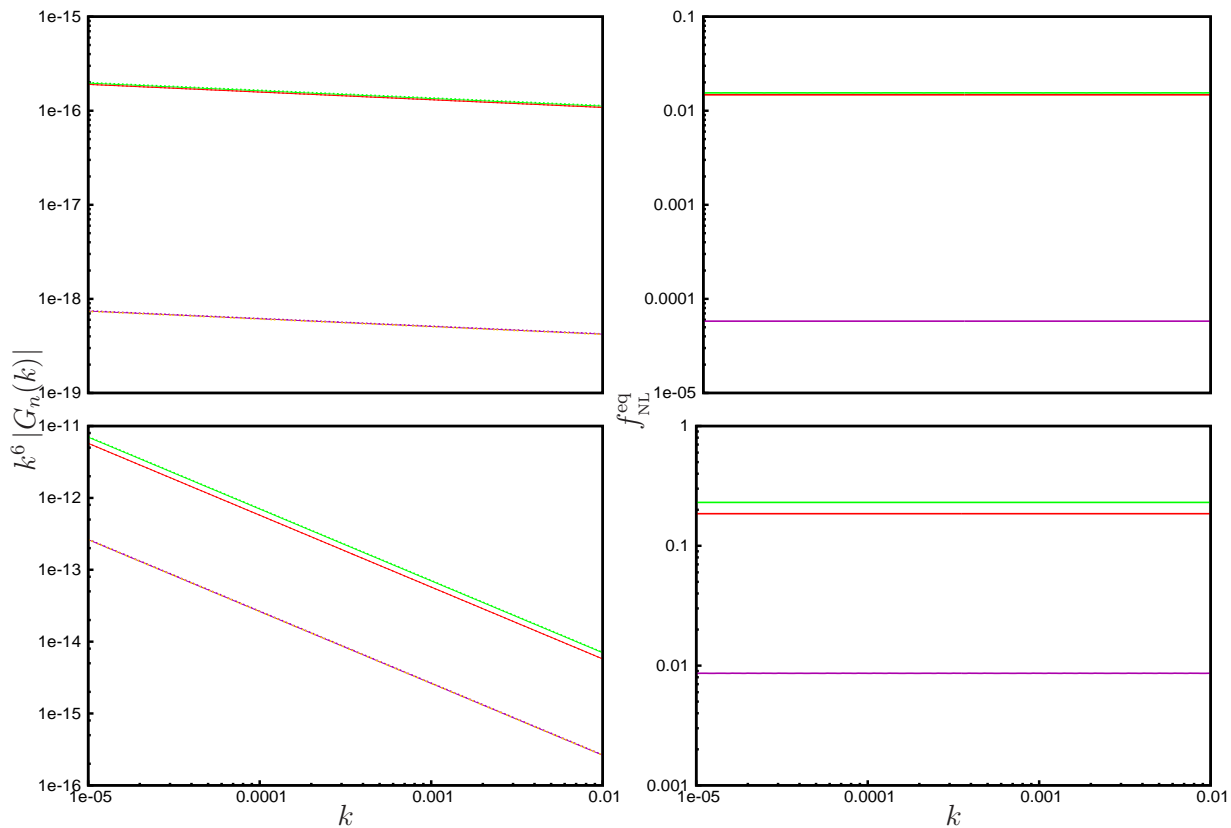}}
\end{center}
\vskip -10pt
\caption{\label{fig:pl} The quantities $k^6$ times the absolute values
of the non-zero contributions in the power law case, viz. $G_1+G_3$, $G_2$
and $G_5+G_6$, obtained numerically, have been plotted on the left for two 
different values of $\gamma$ ($\gamma=-2.02$ on top and $\gamma=-2.25$ 
below), as solid curves with the same choice of colors to represent the 
different quantities as in the previous two figures.
The dots on these curves are the spectral shape arrived at from the analytical
arguments, with the amplitudes chosen to match the numerical results at a 
specific wavenumber.
The dots of a different color on the solid purple curves represents $G_5+G_6$ 
obtained from its relation to $G_1+G_3$ discussed in the text.
The plots on the right are the non-Gaussianity parameter $\fnl^{\mathrm{eq}}$
associated with the different contributions, arrived at using the numerical
code. 
We should mention that we have also arrived at these results independently
using a Mathematica~\cite{mathematica} code.
Note that, as indicated by the analytical arguments, the quantity 
$\fnl^{\mathrm{eq}}$ corresponding to all the contributions turns 
out to be strictly scale invariant for both values of $\gamma$.}
\end{figure}

Consider the case of power law inflation described by the
scale factor~(\ref{eq:pli}) with $\gamma\leq-2$.
In such a case, as we have seen, $\epsilon_1$ is a constant and, hence,
$\epsilon_2$ and $\epsilon_2'$, which involve derivatives of $\epsilon_1$,
reduce to zero. 
Since the contributions due to the fourth and the seventh terms, viz.
$G_4(k)$ and $G_7(k)$, depend on $\epsilon_2'$ and $\epsilon_2$,
respectively [cf. Eqs.~(\ref{eq:cG4}) and~(\ref{eq:G7})], these terms
vanish identically in power law inflation.
As is well known, one can express the modes $f_\vk$ as $v_\vk/z$, where 
$v_\vk$ is the Mukhanov-Sasaki variable.
In power law inflation, one finds that, the variable $v_\vk$ depends only 
on the combination $k\, \eta$ (see, for example, Ref.~\cite{hazra-2012}).
Moreover, since $\epsilon_1$ is a constant in power law expansion, we have 
$f_\vk\propto v_\vk/a$.
Under these conditions, in the equilateral limit, with a simple rescaling 
of the variable of integration in the expressions~(\ref{eq:cG1}), (\ref{eq:cG2}), 
(\ref{eq:cG3}), (\ref{eq:cG5}) and~(\ref{eq:cG6}), it is straightforward to 
show that the quantities $\cG_1$, $\cG_2$, $\cG_3$, $\cG_5$ and $\cG_6$, all 
depend on the wavenumber as $k^{\gamma+1/2}$.
Then, upon making use of the asymptotic form of the modes $f_\vk$, it is
easy to illustrate that the corresponding contributions to the 
bi-spectrum, viz. $G_1+G_3$, $G_2$ and $G_5+G_6$, all behave as
$k^{2\,(2\,\gamma+1)}$.
Since the power spectrum in power law inflation is known to have the form
$k^{2\,(\gamma+2)}$ (see, for example, Refs.~\cite{s-R-sra}), the 
expression~(\ref{eq:fnl-eq}) for $\fnl^{\mathrm{eq}}$ then immediately 
suggests that the quantity will be strictly scale invariant for all $\gamma$. 
In fact, apart from these results, it is also simple to establish the
following relation between the different contributions: $G_5+G_6
=-(3\,\epsilon_1/16)\, (G_1+G_3)$, a result, which, in fact, also holds
in slow roll inflation~\cite{martin-2012a}.
In other words, in power law inflation, it is possible to arrive at the 
spectral dependence of the non-zero contributions to the bi-spectrum
without having to explicitly calculate the integrals involved.
Further, one can establish that the non-Gaussianity parameter 
$\fnl^{\mathrm{eq}}$ is exactly scale independent for any value of $\gamma$.
While these arguments do not help us in determining the amplitude of
the various contributions to the bi-spectrum or the non-Gaussianity
parameter, their spectral shape and the relative magnitude of the 
above-mentioned terms provide crucial analytical results to crosscheck 
our numerical code.    
In Fig.~\ref{fig:pl}, we have plotted the different non-zero contributions
to the bi-spectrum computed using our numerical code and the spectral
dependence we have arrived at above analytically for two different values
of~$\gamma$ in the case of power law inflation.
We have also indicated the relative magnitude of the first and the third
and the fifth and the sixth terms arrived at numerically. 
Lastly, we have also illustrated the scale independent behavior of the 
non-Gaussianity parameter $\fnl^{\mathrm{eq}}$ for both the values of 
$\gamma$.
It is clear from the figure that the numerical results agree well with 
the results and conclusions that we arrived at above analytically.

\subsection{Comparison for an arbitrary triangular configuration}

Let us now turn to the example of the conventional quadratic potential. 
The bi-spectrum in such a case can be evaluated analytically in the slow 
roll approximation for an arbitrary triangular configuration of the
wavevectors~\cite{maldacena-2003,ng-ncsf,ng-reviews}.
In the slow roll limit, one finds that, the contributions due to the 
first, the second, the third and the seventh terms are of the same 
order, while the remaining terms turn out to be comparatively insignificant.
For convenience, let us explicitly write down the contributions to the 
bi-spectrum due to the first, the second, the third and the seventh terms 
arrived at in the slow roll approximation. 
They are given by
\begin{eqnarray}
G_1(\vka,\vkb,\vkc)
&=&\f{H_{_{\rm I}}^4}{16\, \Mpl^4\, \epsilon_1}\,
\l(\f{1}{k_1\, k_2\,k_3}\r)^3\nn\\
& &\times\,\l[\l(1+\f{k_1}{k_{_{\rm T}}}\r)\, 
\f{k_2^{2}\, k_3^{2}}{k_{_{\rm T}}}+{\rm two~permutations}\r],\label{eq:G1-sr}\\
G_2(\vka,\vkb,\vkc)
&=&\f{H_{_{\rm I}}^4}{16\, \Mpl^4\, \epsilon_1}\,
\l(\f{1}{k_1\, k_2\,k_3}\r)^3\,\l(\vka\cdot\vkb+\rm two~permutations\r)\nn\\
& &\times\,
\l[-k_{_{\rm T}}+\f{1}{k_{_{\rm T}}}\, 
\l(k_1\, k_2+k_1\, k_3+k_2\, k_3\r)
+\f{k_1\,k_2\, k_3}{k_{_{\rm T}}^{2}}\r],\\
G_3(\vka,\vkb,\vkc)
&=&-\f{H_{_{\rm I}}^4}{16\, \Mpl^4\, \epsilon_1}\,
\l(\f{1}{k_1\, k_2\,k_3}\r)^3\,
\Biggl[\l(\vka\cdot\vkb\r)\, \f{k_3^2}{k_{_{\rm T}}}\,
\l(2+\f{k_1+k_2}{k_{_{\rm T}}}\r)\nn\\
& &+\,{\rm two~permutations}\Biggr],\\
G_7(\vka,\vkb,\vkc)
&=&2\, \pi^{4}\, \epsilon_2\,\l(\f{1}{k_1\, k_2\,k_3}\r)^3\nn\\
& &\times\,
\l[k_1^{3}\, {\cal P}_{_{\rm S}}(k_2)\,{\cal P}_{_{\rm S}}(k_3)
+\rm two~permutations\r],\label{eq:G7-sr}
\end{eqnarray}
respectively, with $k_{_{\rm T}}=k_1+k_2+k_3$.
The quantities $\epsilon_1$ and $\epsilon_2$ are the first two slow roll
parameters which are assumed to be largely constant, while $H_{_{\rm I}}$ 
denotes the Hubble scale during slow roll inflation.
The scalar power spectrum during slow roll can be written 
as~\cite{texts,reviews}
\begin{equation}
{\cal P}_{_{\rm S}}(k)= 
\f{H_{_{\rm I}}^2}{8\, \pi^2\,\Mpl^2\, \epsilon_1}\, 
\l(\f{k}{k_{\ast}}\r)^{n_{_{\rm S}}-1},
\end{equation}
where $n_{_{\rm S}}=1-2\,\epsilon_1-\epsilon_2$ and $k_{\ast}$ denotes a suitable, 
so-called, pivot scale.

\begin{figure}[!t]
\begin{center}
\resizebox{400pt}{300pt}{\includegraphics{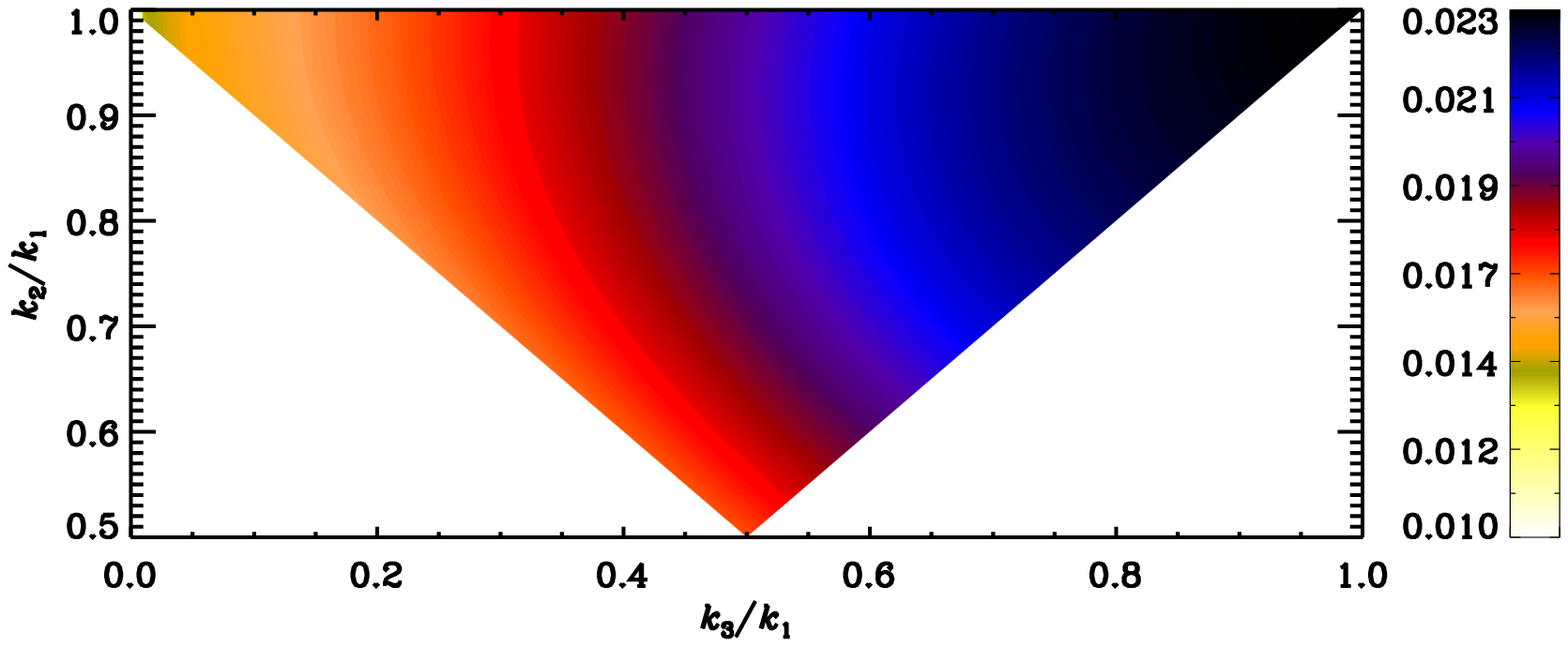}}
\vskip -100pt
\resizebox{400pt}{300pt}{\includegraphics{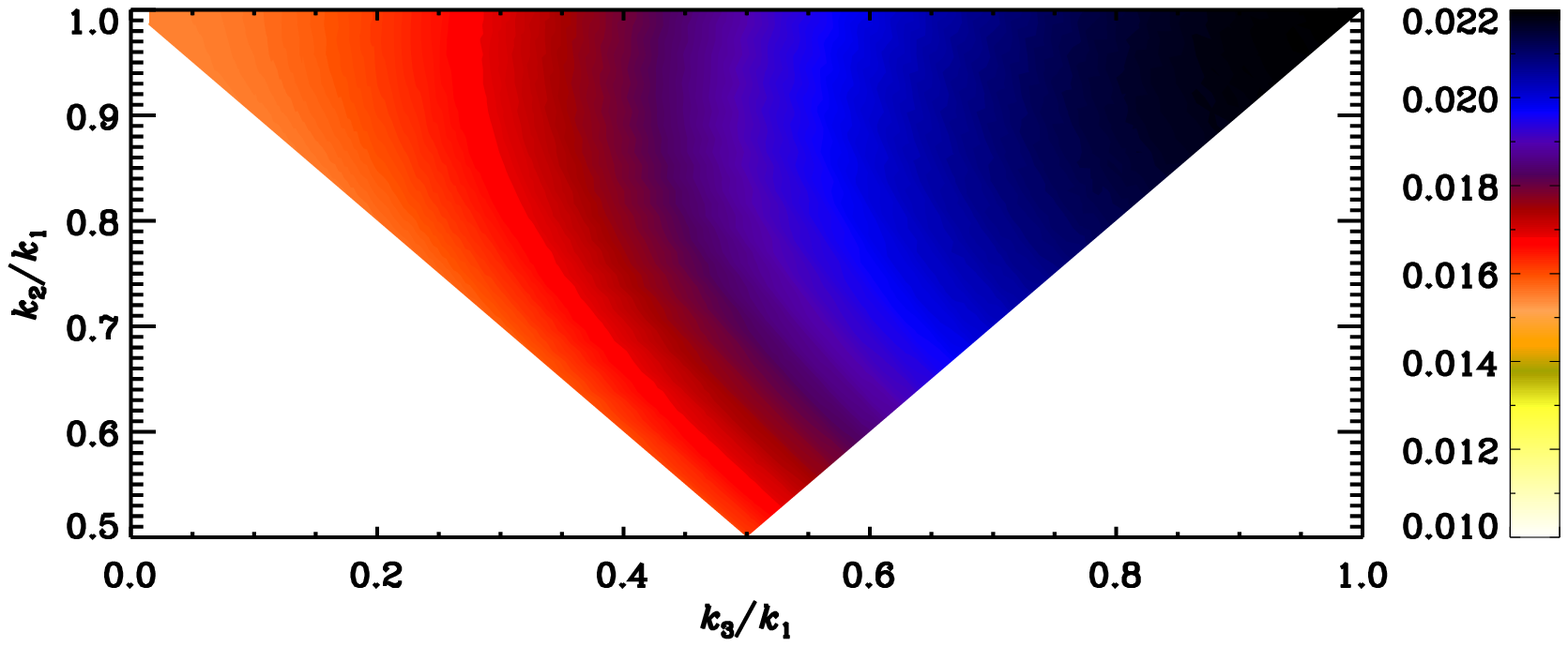}}
\end{center}
\vskip -50pt
\caption{\label{fig:qp-ac}The non-Gaussianity parameter parameter~$\fnl$ 
arising in the case of inflation driven by the archetypical quadratic
potential has been plotted for an arbitrary triangular configuration of the 
wavevectors.
The figure on top corresponds to the analytical expression that has been
arrived at using the slow roll approximation, whereas the figure below 
represents the numerical result from BINGO.
While the analytical result represents the sum of the contributions 
due to the first, the second, the third and the seventh terms [cf. 
Eqs.~(\ref{eq:G1-sr})--(\ref{eq:G7-sr})], the numerical result
takes into account the contributions due to {\it all}\/ the seven terms.
The two figures are strikingly similar.
It should be mentioned here that the ranges of dimensionless ratios of the
wavenumbers $k_3/k_1$ and $k_2/k_1$ for which the parameter~$\fnl$ have been 
plotted is sufficient to reveal the complete structure (in this context, see, 
for instance, Fig.~3 of the last reference in Refs.~\cite{ng-da-reviews}).
Also, note that we have set $k_1=k_{\ast}$ in arriving at the 
above figures.}
\end{figure}
\begin{figure}[!t]
\begin{center}
\resizebox{400pt}{300pt}{\includegraphics{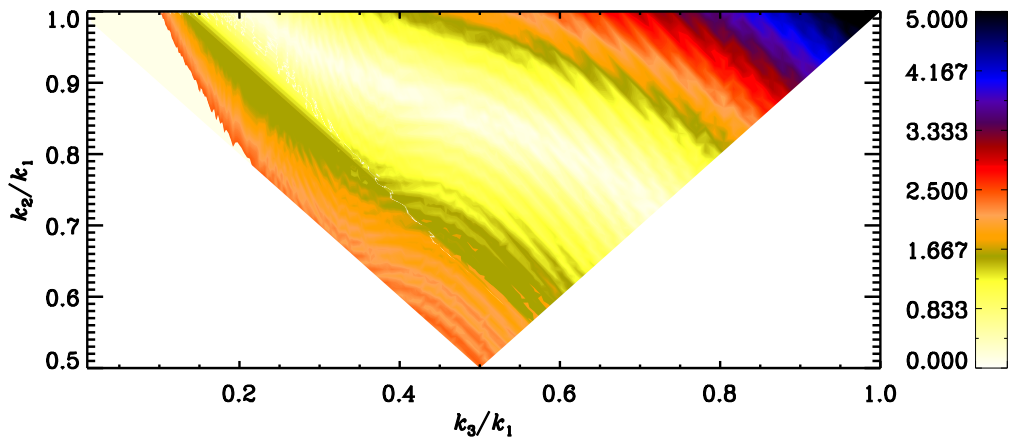}}
\end{center}
\vskip -50pt
\caption{\label{fig:qp-ac-c}The difference in percentage between the 
analytical and the numerical results in the previous figure.
Clearly, the maximum difference is about $5\%$, indicating a rather
good agreement.
Needless to add, BINGO seems to be performing rather well.}
\end{figure}

\par

From the above expressions and, upon using the fact that $\vka+\vkb+\vkc$
has to be zero, one can arrive at the corresponding non-Gaussianity 
parameter $\fnl$ for an arbitrary triangular configuation of the wavevectors.
We find that it can be written as
\begin{eqnarray}
\fnl(\vka,\vkb,\vkc) 
&=& -\f{5}{6}\, \l(\f{k_1}{k_\ast}\r)^{-2\, \l(n_{_{\rm S}}-1\r)}\,
\Biggl[\l(\frac{k_2}{k_1}\r)^{n_{_{\rm S}}-1}\,
\l(\frac{k_3}{k_1}\r)^{n_{_{\rm S}}-1}\nn\\
& &+\l(\f{k_2}{k_1}\r)^3\,\l(\f{k_3}{k_1}\r)^{n_{_{\rm S}}-1}
+\l(\f{k_3}{k_1}\right)^3\,
\l(\f{k_2}{k_1}\r)^{n_{_{\rm S}}-1}\Biggr]^{-1}\nn\\
& &\times\,\Biggl[\epsilon_1\,{\cal F}_{1}(\vka,\vkb,\vkc)
+\f{\epsilon_1}{2}\,{\cal F}_{2}(\vka,\vkb,\vkc)\nn\\
& &-\,\f{\epsilon_1}{2}\,{\cal F}_{3}(\vka,\vkb,\vkc)
+\frac{\epsilon_2}{2}\,{\cal F}_{7}(\vka,\vkb,\vkc)\Biggr], 
\end{eqnarray}
with
\begin{eqnarray}
{\cal F}_{1}(\vka,\vkb,\vkc) 
&=&\l(1+\f{k_2}{k_1}+\f{k_3}{k_1}\r)^{-1}\nn\\ 
& &\times\,\Biggl\{\l(\frac{k_2}{k_1}\r)^2
+\l(\frac{k_3}{k_1}\r)^2
+\l(\f{k_2}{k_1}\right)^2\,\l(\f{k_3}{k_1}\right)^2\nn\\
& &+\,\l(1+\f{k_2}{k_1}+\f{k_3}{k_1}\r)^{-1}\nn\\
& &\times\,\l[\f{k_2}{k_1}\,\l(\f{k_3}{k_1}\r)^2
+\f{k_3}{k_1}\,\l(\f{k_2}{k_1}\r)^2+\l(\f{k_2}{k_1}\r)^2\,
\l(\f{k_3}{k_1}\r)^2\r]\Biggr\},\\
{\cal F}_{2}(\vka,\vkb,\vkc)  
&=&\l[1+\l(\frac{k_2}{k_1}\r)^2+\l(\f{k_3}{k_1}\r)^2\r]\nn\\ 
& &\times\,\Biggl[1+\f{k_2}{k_1}+\f{k_3}{k_1}\nn\\
& &-\,\l(1+\f{k_2}{k_1}+\f{k_3}{k_1}\r)^{-1}\,
\l(\f{k_2}{k_1}+\f{k_3}{k_1}+\f{k_2}{k_1}\,\f{k_3}{k_1}\r)\nn\\ 
& &-\l(1+\f{k_2}{k_1}+\frac{k_3}{k_1}\r)^{-2}\,
\f{k_2}{k_1}\,\f{k_3}{k_1}\Biggr],\\
{\cal F}_{3}(\vka,\vkb,\vkc) 
&=& \l(1+\frac{k_2}{k_1}+\f{k_3}{k_1}\r)^{-1}\nn\\
& &\times\,\Biggl\{\l[2+\l(1+\f{k_2}{k_1}+\f{k_3}{k_1}\r)^{-1}\,
\l(\f{k_2}{k_1}+\f{k_3}{k_1}\r)\r]\nn\\ 
& &\times\,\l[1-\l(\f{k_2}{k_1}\r)^2
-\l(\frac{k_3}{k_1}\r)^2\r]\nn\\
& &-\,\l[2+\l(1+\f{k_2}{k_1}+\f{k_3}{k_1}\r)^{-1}\,
\l(1+\f{k_2}{k_1}\r)\r]\nn\\ 
& &\times\,\l[1+\l(\f{k_2}{k_1}\r)^2
-\l(\frac{k_3}{k_1}\r)^2\r]\,\l(\f{k_3}{k_1}\r)^2\nn\\
& &-\,\l[2+\l(1+\f{k_2}{k_1}+\f{k_3}{k_1}\r)^{-1}\,
\l(1+\f{k_3}{k_1}\r)\r]\nn\\ 
& &\times\,\l[1-\l(\f{k_2}{k_1}\r)^2
+\l(\frac{k_3}{k_1}\r)^2\r]\,\l(\f{k_2}{k_1}\r)^2\Biggr\},\\ 
{\cal F}_{7}(\vka,\vkb,\vkc) 
&=& \l(\f{k_1}{k_\ast}\r)^{2\, \l(n_{_{\rm S}}-1\r)}\,
\Biggl[\l(\frac{k_2}{k_1}\r)^{n_{_{\rm S}}-1}\,
\l(\frac{k_3}{k_1}\r)^{n_{_{\rm S}}-1}\nn\\
& &+\l(\f{k_2}{k_1}\r)^3\,\l(\f{k_3}{k_1}\r)^{n_{_{\rm S}}-1}
+\l(\f{k_3}{k_1}\right)^3\,
\l(\f{k_2}{k_1}\r)^{n_{_{\rm S}}-1}\Biggr].
\end{eqnarray}
In Fig.~\ref{fig:qp-ac}, we have plotted the above analytical 
expression for $\fnl$ in the case of the quadratic potential as well 
as the corresponding numerical result from BINGO for an arbitrary 
triangular configuration of the wavevectors~\cite{ng-da-reviews}.
The two plots look remarkably alike. 
Also, in Fig.~\ref{fig:qp-ac-c}, we have plotted the percentage difference 
between the analytical and the numerical results.
The figure suggests that the analytical and the numerical results match 
within $5\%$.
It is worth stressing that part of the difference can be attributed to the 
slow roll approximation utilized in arriving at the analytical result.
The extent of the agreement reflects the fact that BINGO performs rather
well.
However, we should stress here that the level of accuracy
of BINGO can be improved by essentially carrying out the integrals from a
larger initial value of $k/(a\, H)$ (i.e. from an earlier point in time) and 
simultaneously working with a smaller value of the cut off parameter~$\kappa$.


\subsection{Comparison in the case of the Starobinsky model}

The last of the examples that we shall discuss is the Starobinsky model 
which involves a linear potential with a sharp change in its slope 
[cf.~Eq.~(\ref{eq:p-sm})]. 
In the Starobinsky model, under certain conditions, the complete scalar
bi-spectrum can be evaluated analytically in the equilateral 
limit~\cite{martin-2012a,arroja-2011-2012}.

\par 

\begin{figure}[!tbp]
\begin{center}
\psfrag{allG}[0][1][1.0]{$k^6\, \vert G_{n}(k)\vert$} 
\psfrag{kmpc}[0][1][1.0]{$k/k_0$}
\resizebox{420pt}{280pt}{\includegraphics{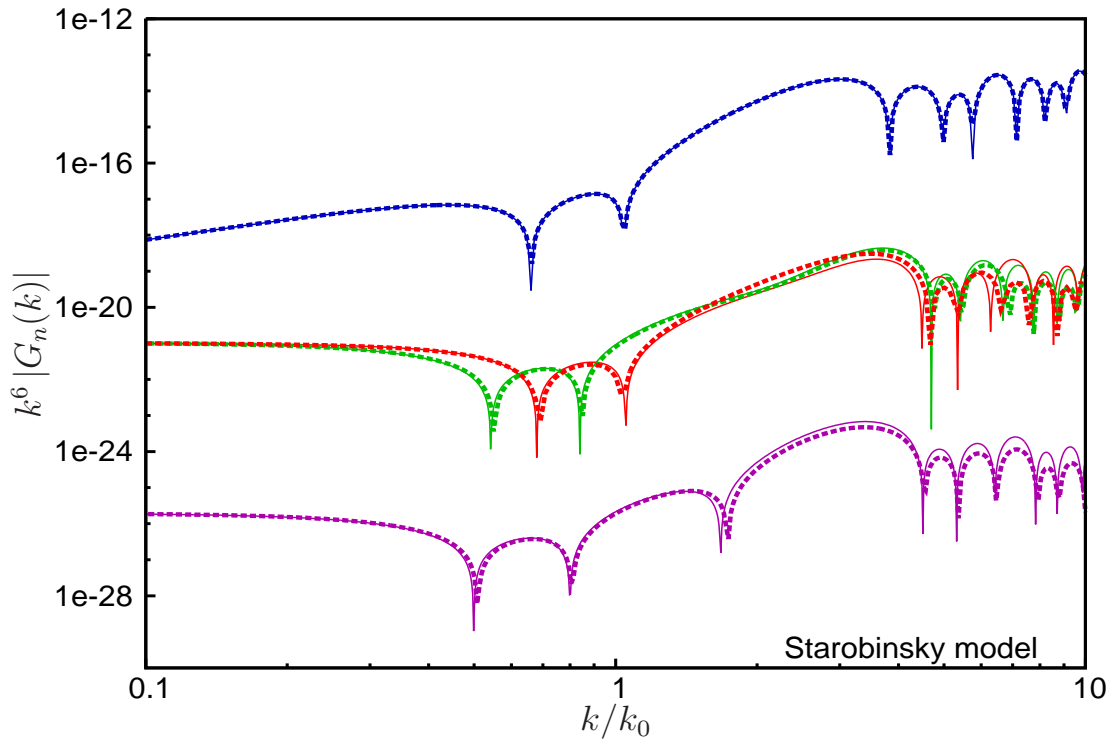}}
\end{center}
\caption{\label{fig:sm-all} The quantities $k^6$ times the 
absolute values of $G_1+G_3$ (in green), $G_2$ (in red), $G_4+G_7$ (in 
blue) and $G_5+G_6$ (in purple) have been plotted as a function of $k/k_0$ 
for the Starobinsky model.
These plots correspond to the following values of the model parameters:
$V_0=2.36\times 10^{-12}\, \Mpl^4$, $A_+=3.35\times 10^{-14}\, \Mpl^3$, 
$A_-=7.26\times 10^{-15}\, \Mpl^3$ and $\phi_0=0.707\, \Mpl$.
Note that $k_0$ is the wavenumber which leaves the Hubble radius when the
scalar field crosses the break in the potential at $\phi_0$.
The solid curves represent the analytical expressions that have been obtained
recently~\cite{martin-2012a,arroja-2011-2012}, while the dashed curves denote
the numerical results computed using our Fortran code.
We find that the numerical results match the analytical results exceptionally 
well (they differ by less than $1\%$) in the case of the 
crucial, dominant contribution to the $\fnl$, viz. due to $G_4+G_7$.}
\end{figure}

As we shall discuss in the next section, in the Starobinsky model, the 
change in the slope causes a brief period of fast roll, which leads to 
sharp features in the scalar power spectrum (see Fig.~\ref{fig:sps-all}). 
It was known that, for certain range of parameters, one could evaluate 
the scalar power spectrum analytically in the Starobinsky model, which 
matches the actual, numerically computed spectrum exceptionally 
well~\cite{starobinsky-1992,martin-2012a}.
Interestingly, it has been recently shown that, in the equilateral limit, 
the model allows the analytic evaluation of the complete scalar bi-spectrum 
too~\cite{martin-2012a,arroja-2011-2012}.
In Fig.~\ref{fig:sm-all}, we have plotted the numerical as well as 
the analytical results for the functions $G_1+G_3$, $G_2$, $G_4+G_7$, 
and $G_5+G_6$ for the Starobinsky model.
We have plotted for parameters of the model for which the analytical 
results are considered to be a good approximation~\cite{martin-2012a}.
It is evident from the figure that the numerical results match the 
analytical ones very well.
Importantly, the agreement proves to be excellent in the case of the 
dominant contribution~$G_4+G_7$.
A couple of points concerning concerning the numerical results in the case 
of the Starobinsky model (both in Fig.~\ref{fig:sps-all} wherein we have 
plotted the power spectrum as well as in Fig.~\ref{fig:sm-all} which
contains the bi-spectrum) require some clarification.
The second (and higher) derivatives of the potential~(\ref{eq:p-sm}) evidently 
contain a discontinuity.
These discontinuities needs to be smoothened in order for the problem to be
numerically tractable.
The spectra and the bi-spectra in the Starobinsky model we have illustrated
have been computed with a suitable smoothing of the discontinuity, while at 
the same time retaining a sufficient level of sharpness so that they closely 
correspond to the analytical results that have been arrived 
at~\cite{martin-2012a,arroja-2011-2012}.


\section{The inflationary models of our interest and the resulting power 
spectra}

In the remainder of this paper, as an immediate application, we shall utilize 
BINGO to examine the power of the non-Gaussianity parameter $\fnl$ to help us 
discriminate between inflationary models that lead to similar features at the 
level of the scalar power spectrum.
We shall list the various inflationary models of our interest in this section,
and describe the power spectra that arise in these models.
In the next section, we shall present the essential results and compare the
$\fnl$ generated in the different models.

\par

Broadly, the models that we shall consider can be categorized into three 
classes.
The first class shall involve potentials which admit a relatively mild 
and brief departure from slow roll.
The second class shall contain small but repeated deviations from slow
roll, while the third and the last class shall involve a short but rather
sharp departure from slow roll. 
Let us now briefly outline the different inflationary models that we shall 
consider under these classes and discuss the scalar power spectra that are 
generated by them. 


\subsection{Inflationary potentials with a step}

\begin{figure}[!t]
\begin{center}
\resizebox{400pt}{300pt}{\includegraphics{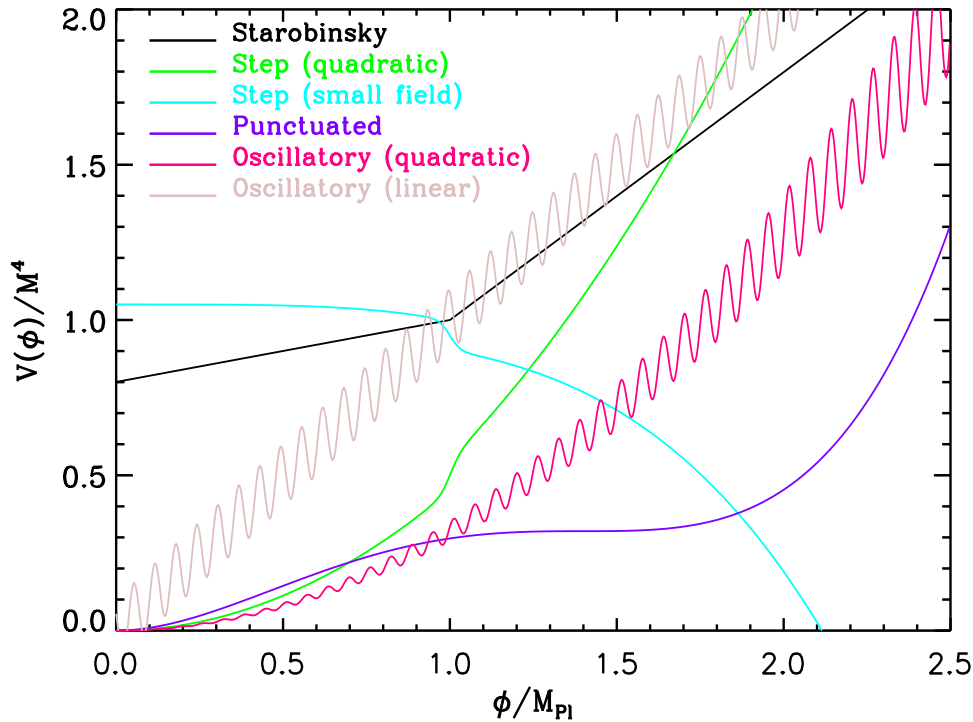}}
\end{center}
\vskip -10pt
\caption{\label{fig:p}An illustration of the potentials in the different 
models of our interest. 
The analytical form of the potentials and the values of the parameters are
discussed in the text.}
\end{figure}

Under the first class, we shall consider models wherein a step has 
been introduced in potentials that otherwise admit only slow roll.
Given a potential, say, $V(\phi)$, we shall introduce the step by 
{\it multiplying}\/ the potential by the following function:
\beq
h_{\mathrm{step}}(\phi)
= 1 + \alpha\, \tanh\l(\frac{\phi-\phi_{0}}{\Delta\phi}\r),
\label{eq:step}
\eeq
as is often done in the literature~\cite{l2240,hazra-2010}.
Clearly, the quantities $\alpha$, $\phi_{0}$ and $\Delta\phi$ denote 
the height, the location and the width of the step, respectively.
We shall consider the effects of the introduction of the step in the 
archetypical quadratic large field model, viz.
\beq
V(\phi) = \frac{1}{2}\,m^2\,\phi^2,\label{eq:qp}
\eeq
where $m$ represents the mass of the inflaton, and a small field 
model governed by the potential
\beq
V(\phi) = V_{0}\, \l[1-\l(\f{\phi}{\mu}\r)^{p}\r].
\label{eq:sfm}
\eeq
We shall specifically focus on the case wherein $p=4$ and $\mu=15\,
\Mpl$, as it leads to a tilt that is consistent with the observations, 
and a smaller tensor-to-scalar ratio (of $r\simeq 0.01$) than the above
quadratic potential\footnote{It may be worth noting that 
such rather large values for the parameter $\mu$ (in Planck units) can be 
considered to be a drawback of the small field model.}.
The introduction of the step has been shown to lead to a short period
of deviation from slow roll inflation, which, in turn, produces a burst 
of oscillations in the scalar power spectrum, resulting in an improved 
fit to the data, in both these models~\cite{hazra-2010}. 
The best fit parameters in the case of the quadratic potential~(\ref{eq:qp}) 
prove to be $m=7.147\times10^{-6}\, \Mpl$, $\alpha= 1.606\times10^{-3}$, 
$\phi_{0}= 14.67\, \Mpl$ and $\Delta\phi= 0.0311\,\Mpl$.
The field is assumed to start on the inflationary attractor at an 
initial value of $\phi_{\mathrm{i}}=16.5\,\Mpl$, so that at least $60$ 
e-folds of inflation takes place.
The small field model~(\ref{eq:sfm}) too is found to lead to a very 
similar spectrum and an almost identical extent of improvement in 
the fit to the data.
The best fit values of the parameters in this case are found to be
$V_0=5.501\times10^{-10}\, \Mpl^4$, $\alpha= -1.569\times10^{-4}$, 
$\phi_{0}=7.888\,\Mpl$ and $\Delta\phi= 9\times 10^{-3}\,\Mpl$.
Further, as in the quadratic case, the field is set to start on the attractor.
The starting value of the field is chosen to be $\phi_{\mathrm{i}}=7.3\,\Mpl$, 
and it rolls through the step towards $\phi\simeq\mu$.
Fig.~\ref{fig:p} contains an illustration of the potentials corresponding
to the quadratic case~(\ref{eq:qp}) and the small field model~(\ref{eq:sfm}) 
in the presence of the step~(\ref{eq:step}) for the values of the parameters 
discussed above.


\subsection{Oscillations in the inflaton potential}\label{subsec:oip}

The second class of models that we shall consider are those which 
lead to small but repeated deviations from slow roll as in the case 
of models which contain oscillatory terms in the inflaton potential.
The first such model that we shall consider is the conventional
quadratic potential~(\ref{eq:qp}) that is modulated by sinusoidal 
oscillations as follows~\cite{pahud-2009,ng-f1,ng-f}:
\beq
V(\phi) = \frac{1}{2}\,m^2\,\phi^2  \l[1 + \alpha\,
\sin\l(\frac{\phi}{\beta}+\delta\r)\r].\label{eq:qpsin}
\eeq
The second model that we shall consider in this context is the axion 
monodromy model which is motivated by string 
theory~\cite{flauger-2010-2011,aich-2011}.
The inflaton potential in such a case is given by
\beq
V(\phi)
=\lambda\, \l[\phi+\alpha\,\cos\l(\frac{\phi}{\beta}+\delta\r)\r].
\label{eq:amm}
\eeq
Evidently, in these potentials, while the parameters~$\alpha$ and
$\beta$ describe the amplitude and the frequency of the superimposed 
oscillations, the parameter~$\delta$ shifts the oscillations within 
one period.
Note that, whereas the amplitude of the oscillation is fixed in the 
axion monodromy model, the amplitude depends quadratically on the 
field in the chaotic model described by the potential~(\ref{eq:qpsin}).
The inflaton oscillates as it rolls down these potentials, and these
oscillations persist all the way until the end of inflation.

\par

The oscillating inflaton leads to small oscillations in the slow roll 
parameters (in this context, see, for instance, Ref.~\cite{hazra-2012b}), 
which, as we shall soon illustrate, results in continuing oscillations in 
the primordial scalar power spectrum.
It is found that, in the case of the potential~(\ref{eq:qpsin}), the
power spectrum corresponding to $m=1.396\times10^{-5}\,\Mpl$, $\alpha=2.56
\times10^{-4}$, $\beta=0.1624\, \Mpl$ and $\delta=2.256$ leads to a modest 
improvement in the fit to the CMB data~\cite{pahud-2009,aich-2011}.
Whereas, the axion monodromy model~(\ref{eq:amm}) leads to a considerable
improvement in the fit for the following values of the parameters
$\lambda=2.513\times10^{-10}\, \Mpl^3$, $\alpha=1.84\times10^{-4}\, \Mpl$,
$\beta=4.50\times10^{-4}\, \Mpl$ and $\delta=0.336$.
While the field is assumed to start on the inflationary attractor at 
the initial value of about $\phi_{\mathrm{i}}=12\,\Mpl$ in the monodromy 
model, the field is evolved from $\phi_{\mathrm{i}} =16\, \Mpl$ in the case 
of the quadratic potential with sinusoidal 
oscillations~\cite{flauger-2010-2011,aich-2011}.
We have plotted the two potentials~(\ref{eq:qpsin}) and~(\ref{eq:amm}) in 
Fig.~\ref{fig:p} for the above values of the parameters.


\subsection{Punctuated inflaton and the Starobinsky model}

We shall consider two models under the last class, both of which are
known to lead to brief but sharp departures from slow roll.
The first of the inflationary models that we shall consider in this
class is described by the following potential containing two parameters 
$m$ and $\lambda$:
\beq
V(\phi) = \frac{m^2}{2}\,\phi^2  
- \frac{\sqrt{2\,\lambda\,(n-1)}\,m}{n}\; \phi^n 
+ \frac{\lambda}{4}\,\phi^{2\,(n-1)}.
\label{eq:mssm-p}
\eeq
The third quantity $n$ that appears in the potential is an integer 
which takes values greater than two. 
Such potentials are known to arise in certain minimal supersymmetric
extensions of the standard model~\cite{pi-motivations}. 
It is worthwhile noting here that the case of $n=3$ has been 
considered much earlier for reasons similar to what we shall 
consider here, viz. towards producing certain features in the 
scalar power spectrum~\cite{hodges-1990}.
In the above potential, the coefficient of the $\phi^n$ term is chosen 
such that the potential contains a point of inflection at, say, $\phi 
= \phi_0$ (i.e. the location where both $\d V/\d\phi$ and $\d^{2}V/
\d\phi^{2}$ vanish), so that $\phi_{0}$ given by
\beq
\phi_{0} 
= \left[\frac{2\,m^2}{(n-1)\,\lambda}\right]^{\frac{1}{2\,(n-2)}}.
\eeq

\par 

If one starts at a suitable value of the field beyond the point the 
inflection in the above potential, it is found that one can achieve 
two epochs of slow roll inflation sandwiching a brief period of 
departure from inflation (lasting for a little less than a e-fold), 
a scenario which has been dubbed as punctuated inflation~\cite{pi}.
In fact, it is the point of inflection, around which the potential 
exhibits a plateau with an extremely small curvature, which permits
such an evolution to be possible.
It is found that the following values for the potential parameters
results in a power spectrum that leads to an improved fit to the CMB 
data in the $n=3$ case: $m=1.5012\times 10^{-7}\,\Mpl$ and $\phi_0=
1.95964\, \Mpl$.
Fig.~\ref{fig:p} contains a plot of the potential~(\ref{eq:mssm-p}) 
corresponding to these values of the parameters.
It should be added that the field is assumed to start from rest at 
an initial value of $\phi_{\mathrm{i}}=11.5\,\Mpl$ on the potential
to arrive at the required behavior.

\par

The second model that we shall consider in the last class is the 
Starobinsky model~\cite{starobinsky-1992}, which we had briefly
discussed in the previous section.
As we shall soon illustrate, the model leads to a scalar power 
spectrum that has a certain resemblance to the spectrum generated 
by punctuated inflation.
The model consists of a linear potential with a sharp change 
in its slope at a given point, and can be described as follows:
\begin{equation}
V(\phi) 
= \l\{\begin{array}{ll}
\displaystyle
V_0 + A_{+}\, \l(\phi-\phi_0\r)\ & {\rm for}\ \phi>\phi_0,\\
\displaystyle
V_0 + A_{-}\, \l(\phi-\phi_0\r)\ & {\rm for}\ \phi<\phi_0.
\end{array}\r.\label{eq:p-sm} 
\end{equation}
We have illustrated this potential in Fig.~\ref{fig:p}.
Evidently, while the value of the scalar field where the slope 
changes abruptly is $\phi_0$, the slope of the potential above 
and below $\phi_{0}$ are given by $A_{+}$ and $A_{-}$, respectively.  
Moreover, the quantity $V_{0}$ denotes the value of the potential at 
$\phi=\phi_0$.   
A crucial assumption of the Starobinsky model is that the value of 
$V_0$ is sufficiently large so that the behavior of the scale factor 
always remains close to that of de Sitter.
The change in the slope causes a short period of deviation from slow 
roll as the field crosses~$\phi_0$.
However, in contrast to the case of the punctuated inflationary scenario, 
where one encounters a brief departure from inflation, inflation 
continues uninterrupted in the Starobinsky model.
We have not compared the Starobinsky model with the data, and we shall
work with two different sets of values for the parameters of the model.
One set corresponds to values of the parameters that allow for the 
comparison of the analytical results for the bi-spectrum that have 
been obtained in this case~\cite{martin-2012a,arroja-2011-2012} with 
the corresponding numerical ones.
The other set shall be chosen to lead to a spectrum that closely 
mimics the power spectrum encountered in punctuated inflation.
In the case of the former, as we had already mentioned in the caption of
Fig.~\ref{fig:sm-all}, we shall choose the following values of
the parameters: $V_0=2.36\times 10^{-12}\, \Mpl^4$, $A_+=3.35\times 
10^{-14}\, \Mpl^3$, $A_-=7.26\times 10^{-15}\, \Mpl^3$ and $\phi_0=0.707\, 
\Mpl$, while, in the case of the latter, we shall work with the same values 
of $A_+$ and $\phi_0$, but shall set $V_0=2.94\times 10^{-13}\, \Mpl^4$, 
and $A_-=3.35\times 10^{-16}\,\Mpl^3$.
Also, we shall work with an initial value of $\phi_{\mathrm{i}}=0.849\, 
\Mpl$ in the first instance and with $\phi_{\mathrm{i}}=1.8\, \Mpl$ in 
the second.
Further, we shall start with field velocities that are determined by the 
slow roll conditions in both the cases. 


\subsection{The power spectra}

We shall now discuss the scalar power spectra generated by the different 
inflationary models that we have listed above. 
We shall highlight here certain aspects of the power spectra that arise 
in the different class of models.

\par

\begin{figure}[!t]
\begin{center}
\psfrag{psk}[0][1][1.0]{${\cal P}_{_{\mathrm{S}}}(k)$} 
\psfrag{kmpc}[0][1][1.0]{$k$}
\resizebox{420pt}{280pt}{\includegraphics{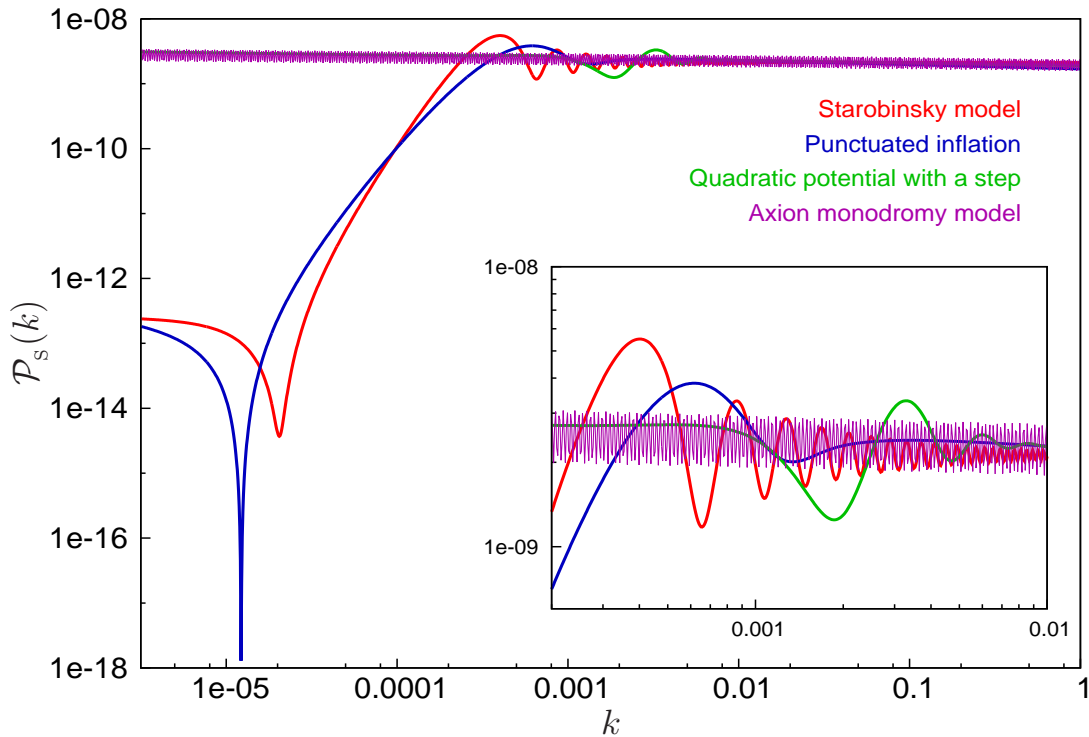}}
\end{center}
\caption{\label{fig:sps-all} The scalar power spectrum in the different 
types of inflationary models that we consider.
The parameters of the Starobinsky model~\cite{starobinsky-1992} have been 
chosen such that the resulting power spectrum closely resembles the 
spectrum that arises in the punctuated inflationary scenario which is 
known to lead to an improved fit to the CMB data~\cite{pi}.
While the models with a step~\cite{l2240,hazra-2010} lead to a burst of 
oscillations over a specific range of scales, inflaton potentials with 
oscillating terms produce modulations over a wide range of scales in the 
power spectrum~\cite{flauger-2010-2011,aich-2011}.
The inset highlights the differences in the various power spectra over a 
smaller range of scales.
We have emphasized the important aspects of these different power spectra 
in some detail in the text.}
\end{figure}

In Fig.~\ref{fig:sps-all}, we have illustrated the scalar power spectrum 
that arises in the different models that we consider.
As we had discussed in Sub-sec.~\ref{subsec:dnm}, the standard Bunch-Davies 
initial conditions are imposed on the perturbations when the modes are well 
inside the Hubble radius.
The modes are evolved on to super Hubble scales, and the spectrum is evaluated
when the amplitude of the curvature perturbation has frozen when they are
sufficiently outside the Hubble radius.  
Let us first focus on the power spectra that arise in the Starobinsky 
model~\cite{starobinsky-1992} and the punctuated inflationary 
scenario~\cite{pi}. 
Note that, both these models lead to a step like feature as well as
a spike in the power spectrum.
The spikes arise due to the sharp departure from slow roll that occurs 
in these models.
While the first slow roll parameter $\epsilon_1$ remains small in the 
Starobinsky model as the field crosses the transition, the second slow 
parameter $\epsilon_2$ turns very large 
briefly~\cite{martin-2012a,arroja-2011-2012}.
In the case of punctuated inflation, $\epsilon_1$ itself grows to a
large value thereby actually interrupting inflation for about a e-fold.
It is this property that results in a sharper spike in the power spectrum
in the case of punctuated inflation than the Starobinsky model.
The overall step in these models is easier to understand, and it simply 
arises due to the difference in the Hubble scales associated with the
slow roll epochs before and after the period of fast roll.  
Both these models also lead to oscillations before the spectra turn
nearly scale invariant on small scales.
The spectra that arises in punctuated inflation, in addition to leading
to a better fit to the outliers at very small multipoles (because of the 
drop in power on these scales), also provides an improvement in the fit
to the outlier at $\ell\simeq 22$~\cite{pi}. 

\par

Let us now turn to the cases of the models with a step in the potential
and the potentials which contain oscillatory terms.
In contrast to the Starobinsky model and the punctuated inflationary
scenario, these models lead to much milder deviations from slow roll.
As a result, they also lead to smaller deviations from a nearly scale
invariant power spectrum.
In the context of models with a step in the potential, a step at the 
correct location and of a suitable amplitude leads to a burst of 
oscillations in the power spectrum which in turn provides an improved 
fit to the outliers at $\ell\simeq 22$ and $40$~\cite{l2240,hazra-2010}.
It is interesting to note in Fig.~\ref{fig:sps-all} that the spectra from
punctuated inflation and the model with a step in the potential match 
briefly as they oscillate near scales corresponding to $\ell\simeq 22$.
Clearly, it is this behavior that leads to a better fit to the outlier 
in the data at these scales.
While all the models that we have discussed on until now result in 
localized features, the models with oscillating inflation potentials are 
somewhat special as they contain non-local features, i.e. they lead to 
continuing oscillations that extend over a wide range of scales in the 
power spectrum.
Needless to add, they arise due to the persisting oscillations of the
scalar field as it rolls up and down the modulations in the potential.
Interestingly, these features provide an improved fit to the data over
a large range of all scales~\cite{flauger-2010-2011,aich-2011}.


\section{Results in the equilateral limit}\label{sec:results} 

We shall now discuss the bi-spectra and the non-Gaussianity parameter $\fnl$
arrived at numerically in the various models of our interest.

\par

\begin{figure}[!t]
\begin{center}
\psfrag{allG}[0][1][1.0]{$k^6\, \vert G_{n}(k)\vert$} 
\psfrag{kmpc}[0][1][1.0]{$k$}
{}\hskip 7pt
\resizebox{342pt}{180pt}{\includegraphics{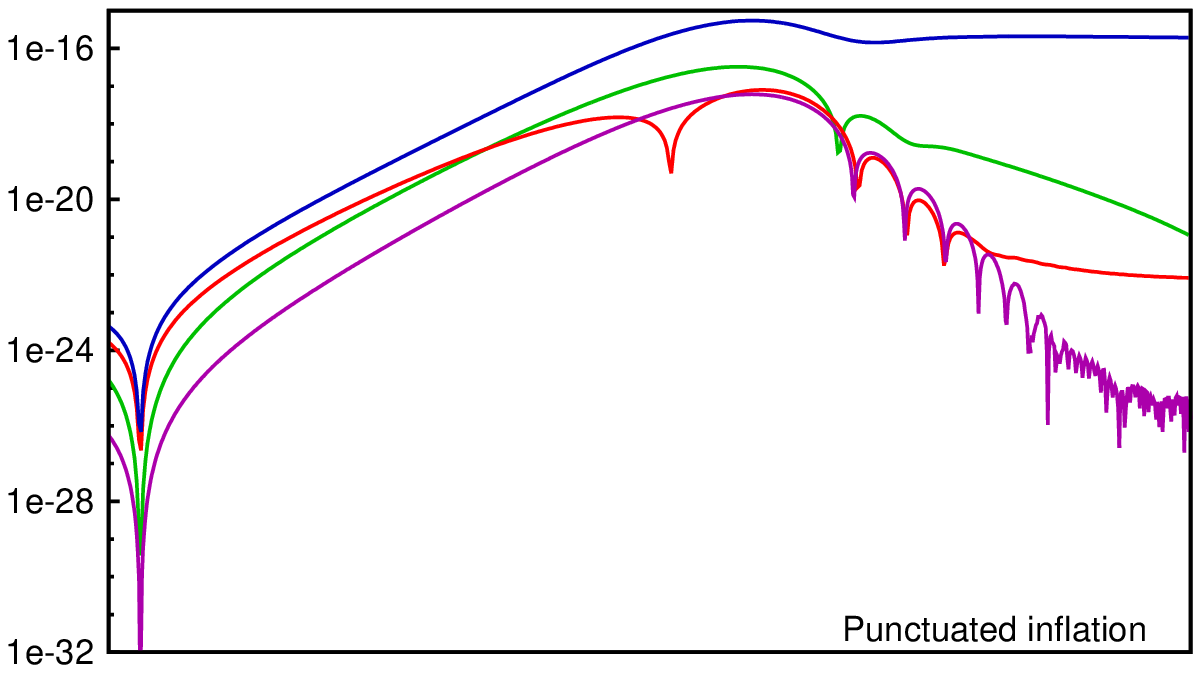}}
\vskip -15pt
\resizebox{350pt}{180pt}{\includegraphics{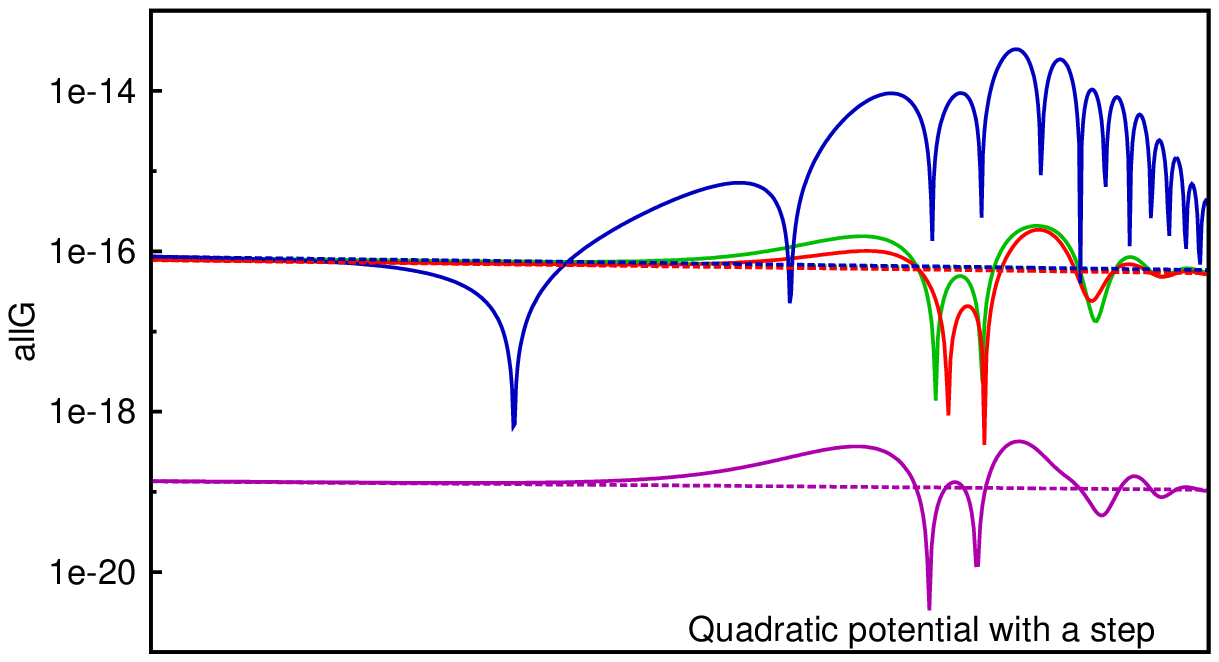}}
\vskip -15pt
\resizebox{350pt}{180pt}{\includegraphics{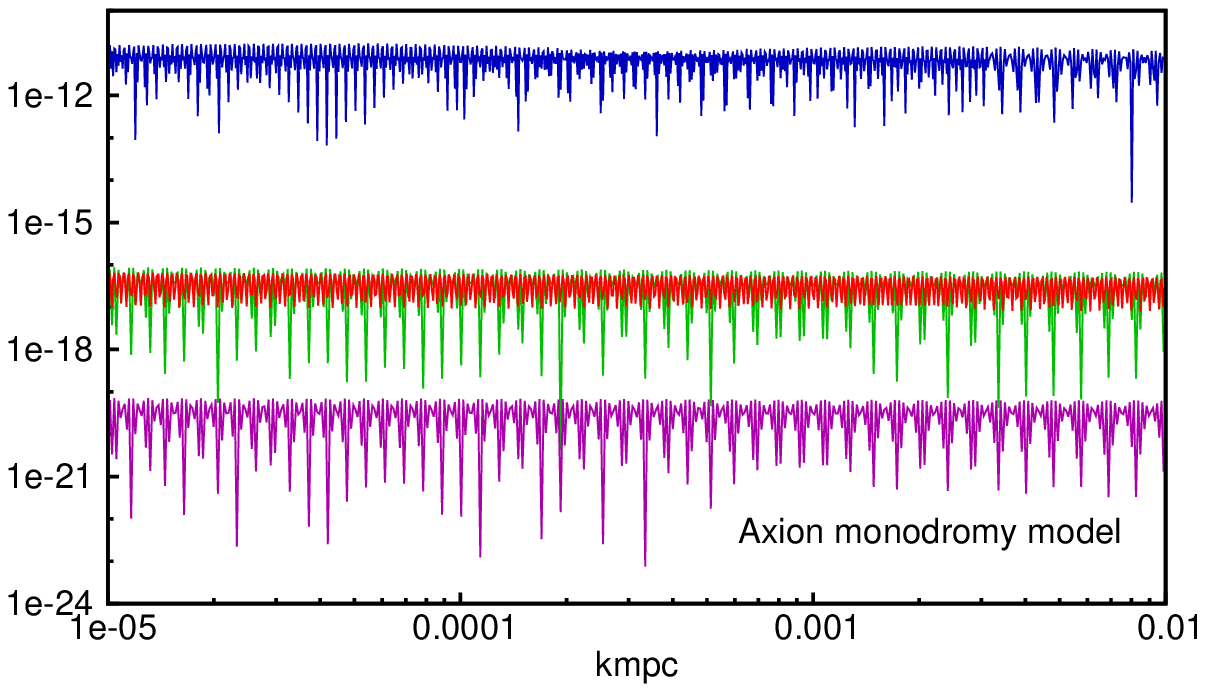}}
\end{center}
\caption{\label{fig:G-all.ps} 
The set of quantities $k^6\, \vert G_{n}(k)\vert$ plotted as in 
Fig.~\ref{fig:sm-all} with the same choice of colors to represent 
the different $G_{n}(k)$. 
The figures on top, in the middle and at the bottom correspond to 
punctuated inflation, the quadratic potential with a step and the 
axion monodromy model, respectively, and they have been plotted 
for values of the parameters that lead to the best fit to the WMAP 
data as discussed in the text~\cite{pi,hazra-2010,aich-2011}.
In the middle figure, the dashed lines correspond to the quadratic
potential when the step is not present.}
\end{figure}

In Fig.~\ref{fig:G-all.ps}, we have plotted the various contributions, 
viz. $G_1+G_3$, $G_2$, $G_4+G_7$ and $G_5+G_6$ (in the equilateral limit) 
for the punctuated inflationary scenario driven by the 
potential~(\ref{eq:mssm-p}), the quadratic potential~(\ref{eq:qp}) with the 
step~(\ref{eq:step}), and the axion monodromy model~(\ref{eq:amm}) which 
contains oscillations in the inflaton potential.
These plots (and also the ones in Fig.~\ref{fig:sm-all}) clearly point 
to the fact that it is the combination $G_4+G_7$ that contributes the
most to the scalar bi-spectrum in these cases~\cite{ng-f1,ng-f}.

\par

\begin{figure}[!t]
\begin{center}
\psfrag{fnl}[0][1][1.0]{$\fnl^{\mathrm{eq}}$} 
\psfrag{absfnl}[0][1][1.0]{$\vert\fnl^{\mathrm{eq}}\vert$} 
\psfrag{kmpc}[0][1][1.0]{$k$}
\resizebox{300pt}{150pt}{\includegraphics{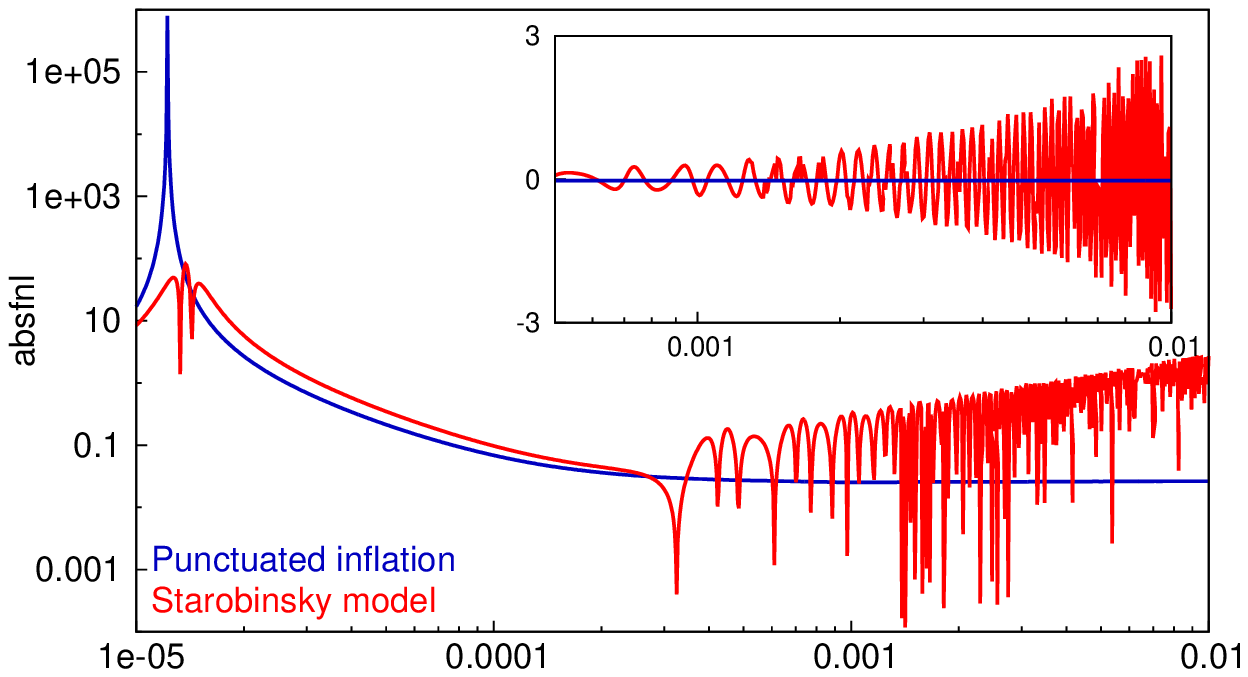}}
\vskip 2pt\hskip 11pt
\resizebox{287pt}{150pt}{\includegraphics{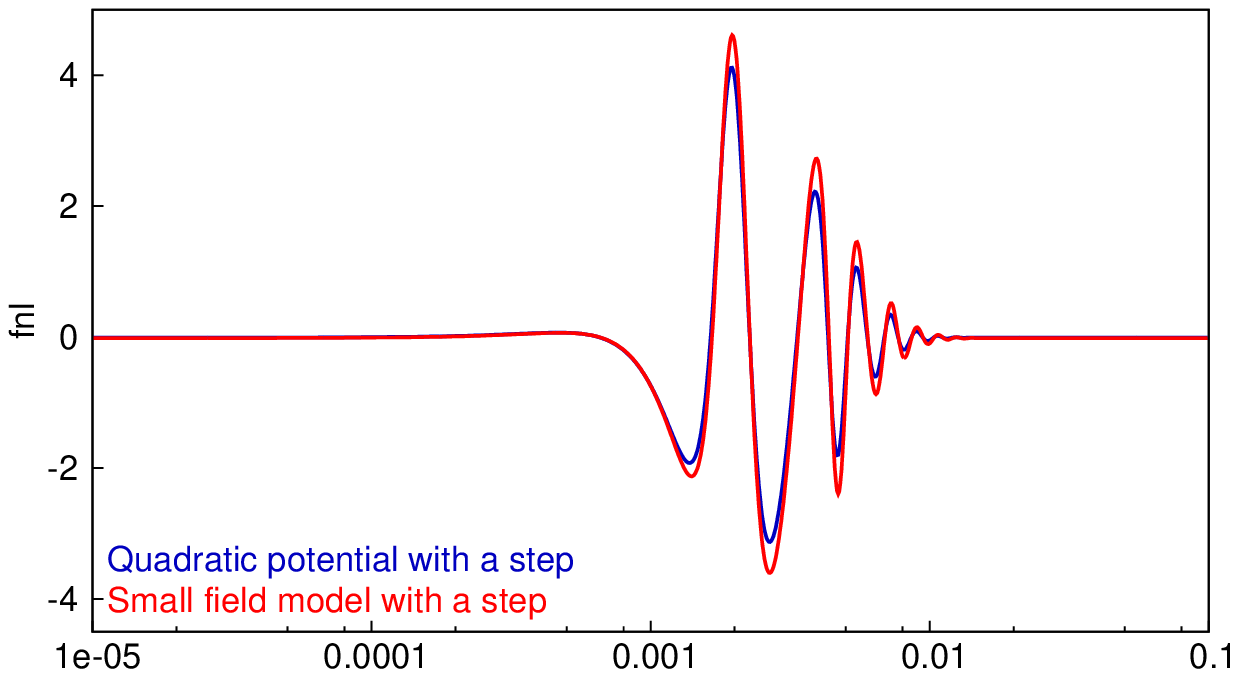}}
\vskip 2pt
\resizebox{300pt}{150pt}{\includegraphics{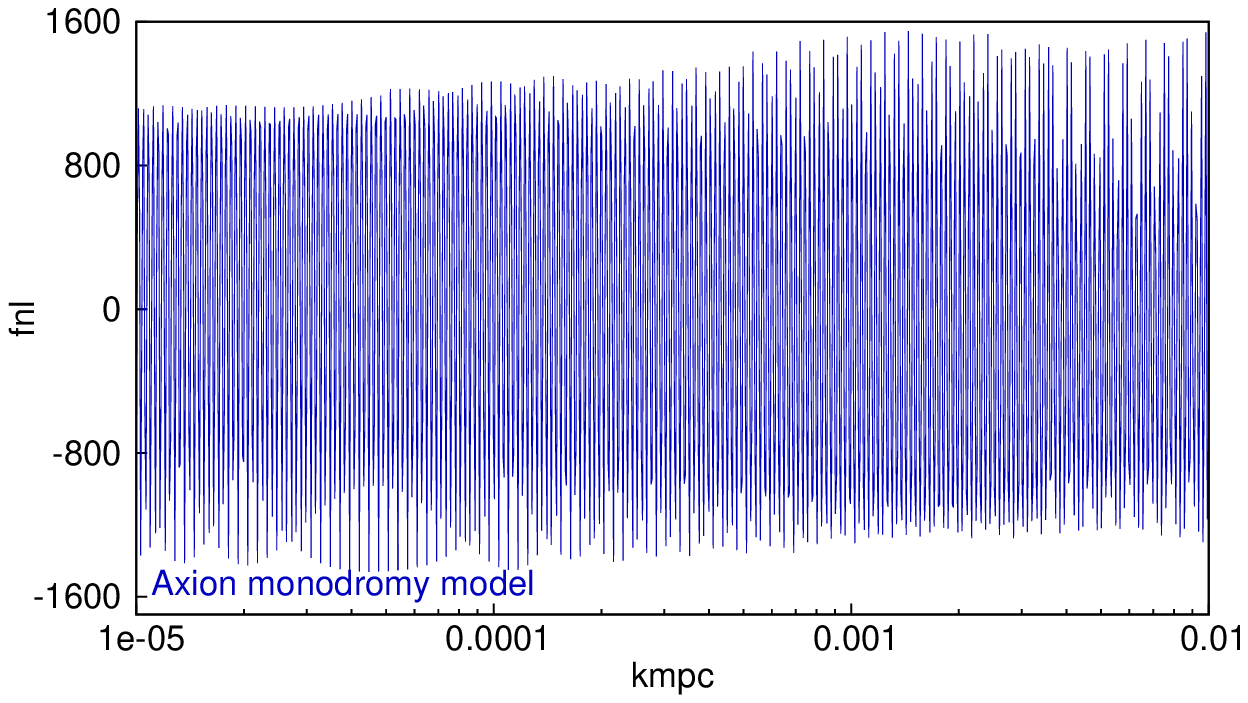}}
\end{center}
\vskip -12pt
\caption{\label{fig:fnl-all}A plot of $\fnl^{\mathrm{eq}}$ 
corresponding to the various models that we have considered.
The figure at the top contains the absolute value of 
$\fnl^{\mathrm{eq}}$, plotted on a logarithmic scale (for convenience 
in illustrating the extremely large values that arise), in the 
Starobinsky model and the punctuated inflationary scenario.
The inset highlights the growth in $\fnl^{\mathrm{eq}}$ 
at large wavenumbers in the case of the Starobinsky model, in conformity 
with the conclusions that have also been arrived at 
analytically~\cite{arroja-2011-2012}.
The figure in the middle contains $\fnl^{\mathrm{eq}}$ for the cases 
wherein a step has been introduced in a quadratic potential and 
a small field model.
The figure at the bottom corresponds to that of the axion monodromy
model.   
As we have described, these sets of models lead to scalar power spectra 
with certain common characteristics.
Needless to say, while $\fnl^{\rm eq}$ is considerably different in the 
first and the last sets of models, it is almost the same in the case of 
models with a step. 
These similarities and differences can be attributed completely to the 
background dynamics.}
\end{figure}

In Fig.~\ref{fig:fnl-all}, we have plotted the quantity $\fnl^{\mathrm{eq}}$ 
due to the dominant contribution that arises in the various models 
that we have considered.
It is clear from this figure that, while in certain cases $\fnl^{\rm eq}$ 
can prove to be a good discriminator, it cannot help in others, and its 
ability to discriminate depends strongly on the differences in the 
background dynamics.
For instance, the evolution of the first two slow roll parameters are very
similar when a step is introduced in either the quadratic potential or a
small field model~\cite{hazra-2010}.
Hence, it is not surprising that the $\fnl^{\mathrm{eq}}$ behaves in a 
similar fashion in both these models.
Whereas, $\fnl^{\mathrm{eq}}$ proves to be substantially different in 
punctuated inflation and the Starobinsky model. 
Recall that, in the Starobinsky model, the first slow roll parameter remains 
small throughout the evolution.
In contrast, it grows above unity for a very short period (leading to 
a brief interruption of the accelerated expansion) in the punctuated 
inflationary scenario.
It is this departure from inflation that leads to a sharp drop in the power 
spectrum and a correspondingly sharp rise in the parameter $\fnl^{\mathrm{eq}}$ 
in punctuated inflation.
In fact, this occurs for modes that leave the Hubble radius just before 
inflation is interrupted~\cite{jain-2007}.
However, note that, $\fnl^{\mathrm{eq}}$ grows with $k$ at large wavenumbers 
in the Starobinsky model.
This can be attributed to the fact that $\epsilon_2'$, which determines
the contribution due to the fourth term, diverges due to the discontinuity 
in the second derivative of the potential~\cite{arroja-2011-2012}.
But, we should hasten to clarify that this can be shown to be an unphysical 
artifact arising due to the presence of a Heaviside function in the 
description of the potential.
Similarly, we find that $\fnl^{\mathrm{eq}}$ is rather large in the axion
monodromy model in contrast to the case wherein the conventional 
quadratic potential is modulated by an oscillatory term. 
The large value of $\fnl^{\rm eq}$ that arises in the monodromy model 
can be attributed to the resonant behavior encountered in the
model~\cite{flauger-2010-2011,ng-f1,ng-f,aich-2011}.
In fact, we have also evaluated the $\fnl^{\mathrm{eq}}$ for the case 
of quadratic potential modulated by sinusoidal oscillations, which too 
leads to continuing, periodic features in the scalar power 
spectrum~\cite{pahud-2009,aich-2011}.
However, we find that the $\fnl^{\mathrm{eq}}$ in such a case proves to 
be rather small (of the order $10^{-2}$ or so).


\section{Discussion}\label{sec:d}

In this work, we have presented BINGO, a code for the efficient computation
of the bi-spectrum and the non-Gaussianity parameter $\fnl$ in single field
inflationary models involving the canonical scalar field.
While there has been previous efforts in the literature towards numerically 
calculating the bi-spectrum and the non-Gaussianity parameter $\fnl$, our
effort in arriving at BINGO can be said to be more complete for the 
following four reasons.
First and foremost, we have explicitly illustrated that the procedures that 
BINGO adopts to evaluate the integrals involved are robust.
On the one hand, we have converged on a suitable form for the cut off
required in the sub-Hubble domain (in particular, the value of the cut off 
parameter $\kappa$) and the corresponding lower limit of the integrals, after 
a careful investigation of their effects on the amplitude of the bi-spectrum.
On the other, we have shown that the super-Hubble contributions to the complete
bi-spectrum are negligible, which allows us to arrive at a convenient upper
limit for the integrals.
Secondly, BINGO can calculate all the contributions to the bi-spectrum.
Thirdly, it can evaluate the bi-spectrum for an arbitrary triangular configuation 
of the wavevectors.
Lastly, we have compared the numerical results from BINGO with the analytical 
expressions available in certain specific cases, an exercise which illustrates 
that BINGO can be accurate to better than $5\%$.
As we had mentioned in the introductory section, we have made a version of BINGO, 
one that focuses on the equilateral limit, available 
online at {\tt https://www.physics.iitm.ac.in/{\textasciitilde{}}sriram/bingo/bingo.html}.
Moreover, if needed, as we have pointed out earlier, the level 
of accuracy of BINGO can be improved by carrying out the integrals from a larger 
initial value of $k/(a\, H)$ (i.e. from an earlier point in time) and 
simultaneously working with a smaller value of the cut off parameter~$\kappa$.
In fact, in the code that is publicly available, these parameters have been
allowed to be set by the user, as is desired.

\par

After presenting BINGO and illustrating the extent of its accuracy, we had 
made use of the code to examine the power of the non-Gaussianity parameter 
$\fnl$ to lift degeneracies between various single field inflationary models 
involving the canonical scalar field at the level of the power spectrum.  
With this goal in mind, using BINGO, we have evaluated the quantity 
$\fnl^{\mathrm{eq}}$ in a slew of models that generate features in 
the scalar perturbation spectrum.
We find that the amplitude of $\fnl^{\mathrm{eq}}$ proves to be rather 
different when the dynamics of the background turns out reasonably 
different, which, in retrospect, need not be surprising at all. 
For instance, models such as the punctuated inflationary scenario and the
Starobinsky model which lead to very sharp features in the power spectrum 
also lead to substantially large $\fnl$. 
Such possibilities can aid us discriminate between the models to some extent. 
We had focused on evaluating the quantity $\fnl$ in the equilateral limit.  
It will be worthwhile to compute the corresponding values in the other limits
as well as for the other forms (such as the squeezed and the orthogonal ones) 
of the non-Gaussianity parameter~\cite{wmap-2012}.
In particular, it will be interesting to examine if the so-called consistency 
relation between the local non-Gaussianity parameter $\fnl$ and the scalar 
spectral index in the squeezed limit is valid even in situations wherein 
extreme deviations from slow roll occur (in this context, 
see Refs.~\cite{ng-f1,ng-f,martin-2012b,creminelli-2004,renaux-petel-2010,chen-2013}).
Actually, since many of the models leading to features
have interesting bi-spectral shapes, it is important that we arrive at
the spectral shapes for the various models for an arbitrary triangular 
configurations of the wavenumbers.
It is also important that, apart from computing the local non-Gaussianity 
parameter as we have done here, one evaluates the corresponding equilateral 
and the orthogonal parameters as well.
We are currently working on extending the scope of BINGO to these cases.

\par

We would like to conclude by highlighting one important point. 
Having computed the primordial bi-spectrum, the next logical step would be 
to compute the corresponding CMB bi-spectrum, an issue which we have not 
touched upon as it is beyond the scope of the current work. 
While tools seem to be available to evaluate the CMB bi-spectrum based 
on the first order brightness function, the contribution due to the 
brightness function at the second order remains to be understood 
satisfactorily (in this context, see Ref.~\cite{pitrou-2010} and the 
last reference in Ref.~\cite{ng-da-reviews}).
This seems to be an important aspect that is worth investigating closer.


\section*{Acknowledgments}

We would like to acknowledge the use of the high performance computing 
facilities at the Harish-Chandra Research Institute, Allahabad, India
({\tt http://cluster.hri.res.in}). 
DKH also wishes to acknowledge support from the Korea Ministry of Education, 
Science and Technology, Gyeongsangbuk-Do and Pohang City for Independent 
Junior Research Groups at the Asia Pacific Center for Theoretical Physics. 

 
\end{document}